\documentclass[10pt,twocolumn,letterpaper]{article}

\usepackage{cvpr}              

\usepackage[dvipsnames]{xcolor}

\definecolor{cvprblue}{rgb}{0.21,0.49,0.74}
\usepackage[pagebackref,breaklinks,colorlinks,citecolor=cvprblue]{hyperref}

\def\confName{CVPR}
\def\confYear{2024}

\usepackage{epsfig}
\usepackage{multicol}
\usepackage{multirow}

\title{SeD: Semantic-Aware Discriminator for Image Super-Resolution}

\author{Bingchen Li\textsuperscript{\rm 1}\footnotemark[1] , Xin Li\textsuperscript{\rm 1}\footnotemark[1] , Hanxin Zhu\textsuperscript{\rm 1}, Yeying Jin\textsuperscript{\rm 2}, Ruoyu Feng\textsuperscript{\rm 1}, Zhizheng Zhang\textsuperscript{\rm 3}, Zhibo Chen\textsuperscript{\rm 1}\footnotemark[2]\\
\textsuperscript{\rm 1}University of Science and Technology of China \\
\textsuperscript{\rm 2}National University of Singapore \quad
\textsuperscript{\rm 3}Galbot \\
\tt\small \{lbc31415926, lixin666, hanxinzhu\}@mail.ustc.edu.cn, jinyeying@u.nus.edu, \\ \tt\small ustcfry@mail.ustc.edu.cn, zhangzz@galbot.com,  chenzhibo@ustc.edu.cn
}

\begin{document}
\maketitle
\renewcommand{\thefootnote}{\fnsymbol{footnote}}
\footnotetext[1]{Equal contribution}
\footnotetext[2]{Corresponding Author}
\begin{abstract}
Generative Adversarial Networks (GANs) have been widely used to recover vivid textures in image super-resolution (SR) tasks. In particular, one discriminator is utilized to enable the SR network to learn the distribution of real-world high-quality images in an adversarial training manner. However, the distribution learning is overly coarse-grained, which is susceptible to virtual textures and causes counter-intuitive generation results. To mitigate this, we propose the simple and effective Semantic-aware Discriminator (denoted as SeD), which encourages the SR network to learn the fine-grained distributions by introducing the semantics of images as a condition. Concretely, we aim to excavate the semantics of images from a well-trained semantic extractor. Under different semantics, the discriminator is able to distinguish the real-fake images individually and adaptively, which guides the SR network to learn the more fine-grained semantic-aware textures. To obtain accurate and abundant semantics, we take full advantage of recently popular pretrained vision models (PVMs) with extensive datasets, and then incorporate its semantic features into the discriminator through a well-designed spatial cross-attention module. In this way, our proposed semantic-aware discriminator empowered the SR network to produce more photo-realistic and pleasing images. Extensive experiments on two typical tasks, \ie, SR and Real SR have demonstrated the effectiveness of our proposed methods. The code will be available at \url{https://github.com/lbc12345/SeD}.
\end{abstract}

\section{Introduction}
Deep learning has accelerated the great development of Single Image Super-resolution (SISR)~\cite{SRCNN,EDSR,ESRGAN,swinir,RealESRGAN,SR3}, which aims to recover the vivid high-resolution (HR) image from its degraded low-resolution (LR) counterpart. In general, there are two typical optimization objectives  for existing SISR methods, \ie, objective and subjective quality. To pursue the higher objective quality, a series of elaborate frameworks based on convolutional neural networks (CNN)~\cite{EDSR,RDN,RCAN,SAN,LDI}, Transformer~\cite{swinir,IPT,TTSR,HNCT,ESRT} have been proposed to improve the representation ability of the SR network through the constrain of the pixel-wise loss function (\eg, L1 loss and MSE loss). However, the pixel-wise loss function cannot enable the SR network to have a promising hallucination capability, which performs poorly on texture generation and results in an unsatisfied subjective quality~\cite{SRGAN,li2023diffusion}.

\begin{figure}
    \centering
    \includegraphics[width=1.\linewidth]
    {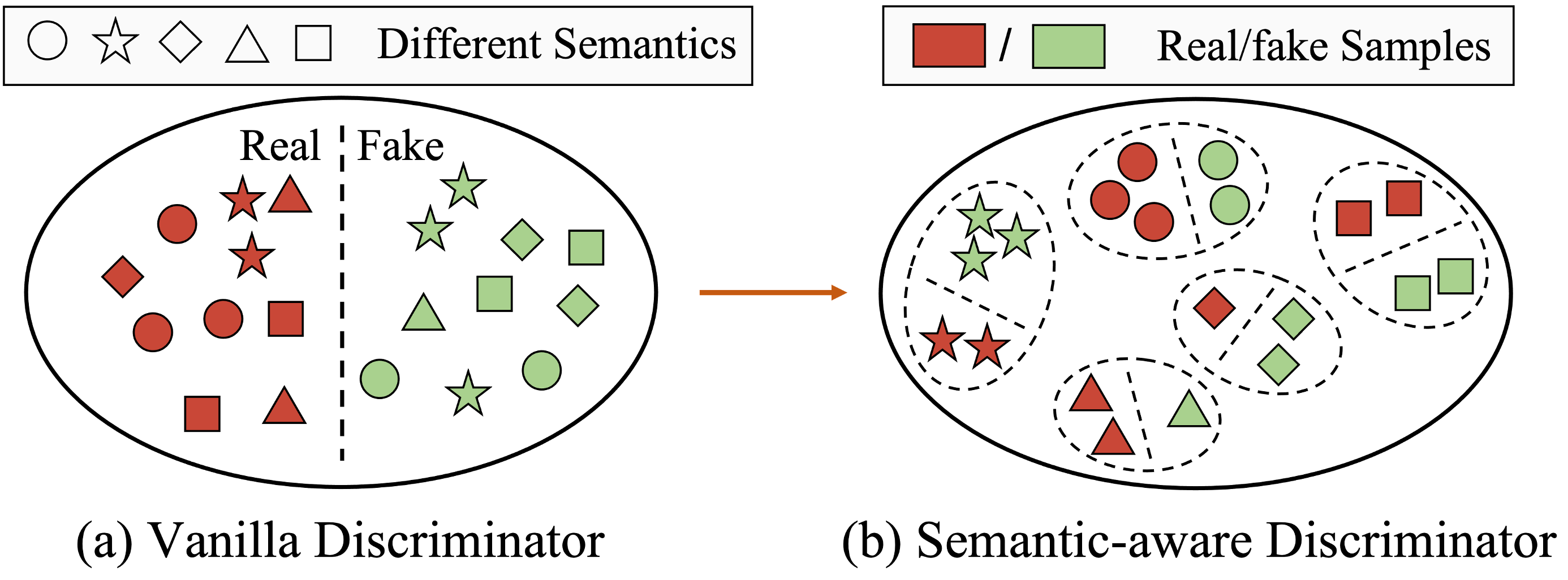}
    \caption{Comparison between the vanilla discriminator and our proposed semantic-aware discriminator.}
    \label{fig:intro}
\end{figure}

To improve the subjective quality, it is necessary to generate human-pleasing textures for distorted images. Inspired by the generative adversarial network (GAN)~\cite{GAN}, a series of pioneering works~\cite{SRGAN,ranksrgan,ESRGAN,BSRGAN,RealESRGAN} regard the SR network as the generator, and then introduce the discriminator to enable the SR networks with realistic texture generation capability. There are three typical discriminators for SR networks, \ie, image-wise discriminator~\cite{VGG,styleganv2}, patch-wise discriminator~\cite{patchgan},  and pixel-wise discriminator~\cite{unet-d}. In particular, the image-wise discriminator aims to distinguish the real/fake images from the global distributions, \eg, VGG-like discriminator in ~\cite{SRGAN,ranksrgan,ESRGAN}. However, the image-wise distribution is so coarse-grained that causes the network to produce non-ideal local textures. To enhance the local textures,~\cite{patchgansr1,patchgansr2,patchgansr3} utilize the patch-wise discriminator to determine the patch-wise distribution and adjust the patch size with different sizes of receptive fields. Furthermore, the pixel-wise discriminator~\cite{unet-d} distinguishes the real/fake distribution in a per-pixel manner, while bringing the large computational cost. However, the above works ignore the fact that the textures of one image should meet the distribution of its semantics. It is necessary to achieve the fine-grained semantic-ware texture generation for SR.

To generate the semantic-aware textures, one intuitive method is to integrate the semantics of images into the generator, which enables the generator the semantic adaptive capability. Actually, this brings two obvious drawbacks to semantic-aware texture generation: 1) The low-quality image may yield worse and even error semantic extraction, which prevents reasonable texture generation. 2) In the inference stage, the complex semantic extractor will cause the catastrophic growth of computation and model complexity for the SR network. In contrast, we aim to achieve the semantic-aware texture generation for SR from another perspective, \ie, Discriminator. 

In this paper, we propose the first Semantic-aware Discriminator for Image Super-resolution (SR), dubbed SeD, which is inspired by ~\cite{zheng2023panoptically}. It is noteworthy that the discriminator works by distinguishing whether the image/patch/pixel is real or fake. With adversarial training, it is enabled to measure the distribution distance between the generated image and its reference image. As shown in Fig.~\ref{fig:intro}, the vanilla discriminator only measures the distribution distance while ignoring the semantics, which is susceptible to the coarse-grained average textures (\eg noise). To mitigate this, we introduce the semantics extracted from recent popular pretrained vision models (PVMs) (\eg ResNet-50 or CLIP) as the condition of the discriminator, which lets the discriminator measure the distribution distance individually and adaptively for different semantics (See Fig.~\ref{fig:intro}(b)). With the constraint of the semantic-aware discriminator, the SR networks (\ie, generator) can achieve more fine-grained semantic-aware texture generation.

There is one crucial step to achieve a semantic-aware discriminator, \ie, how to extract the semantics and incorporate them into a discriminator. A nai\"ve strategy is to directly introduce the features of the last layer from a pretrained classification network as the semantics and concatenate them in a discriminator as a condition. However, it will prevent effective semantic guidance for discriminators, since it does not consider the characteristics of SR. To achieve fine-grained semantic-aware texture generation, one intuition is that the semantics is required to be pixel-wise due to semantics in different regions might be different. Considering this, we extract the semantics from the middle features of the PVMs. Depart from this, we also design one semantic-aware fusion block (SeFB) to better incorporate the semantics into the discriminator. Concretely, in SeFB, we regard the semantics extracted from PVMs as the query, to warp the semantic-aware image feature to the discriminator through cross-attention, which brings better semantic guidance. We validate the effectiveness of our proposed SeD on two typical SR tasks, \ie, classical and Real-world image SR tasks. Our SeD is general and applicable to different benchmarks for GAN-based SR, \eg, ESRGAN~\cite{ESRGAN}, RealESRGAN~\cite{RealESRGAN}, and BSRGAN~\cite{BSRGAN}.

The contributions of this paper can be summarized as follows:
\begin{itemize}
    \item We pinpoint the importance of fine-grained semantic-aware texture generation for SR, and for the first time propose a semantic-aware discriminator (SeD) for the SR task, by incorporating the semantics from pretrained vision models (PVMs) into the discriminator.
    \item To better incorporate the guidance of semantics for the discriminator, we propose the semantic-aware fusion block (SeFB) for SeD, which extracts the pixel-wise semantics and warped the semantic-aware image features into the discriminator through the cross-attention manner. 
    \item Extensive experiments on two typical SR tasks, \ie, classical and Real-world image SR have revealed the effectiveness of our proposed SeD. Moreover, our SeD can be easily integrated into many benchmarks for the GAN-based SR methods in a plug-and-play manner. 
    \end{itemize}

\begin{figure*}[t!]
    \centering
    \includegraphics[width=1.0\linewidth]{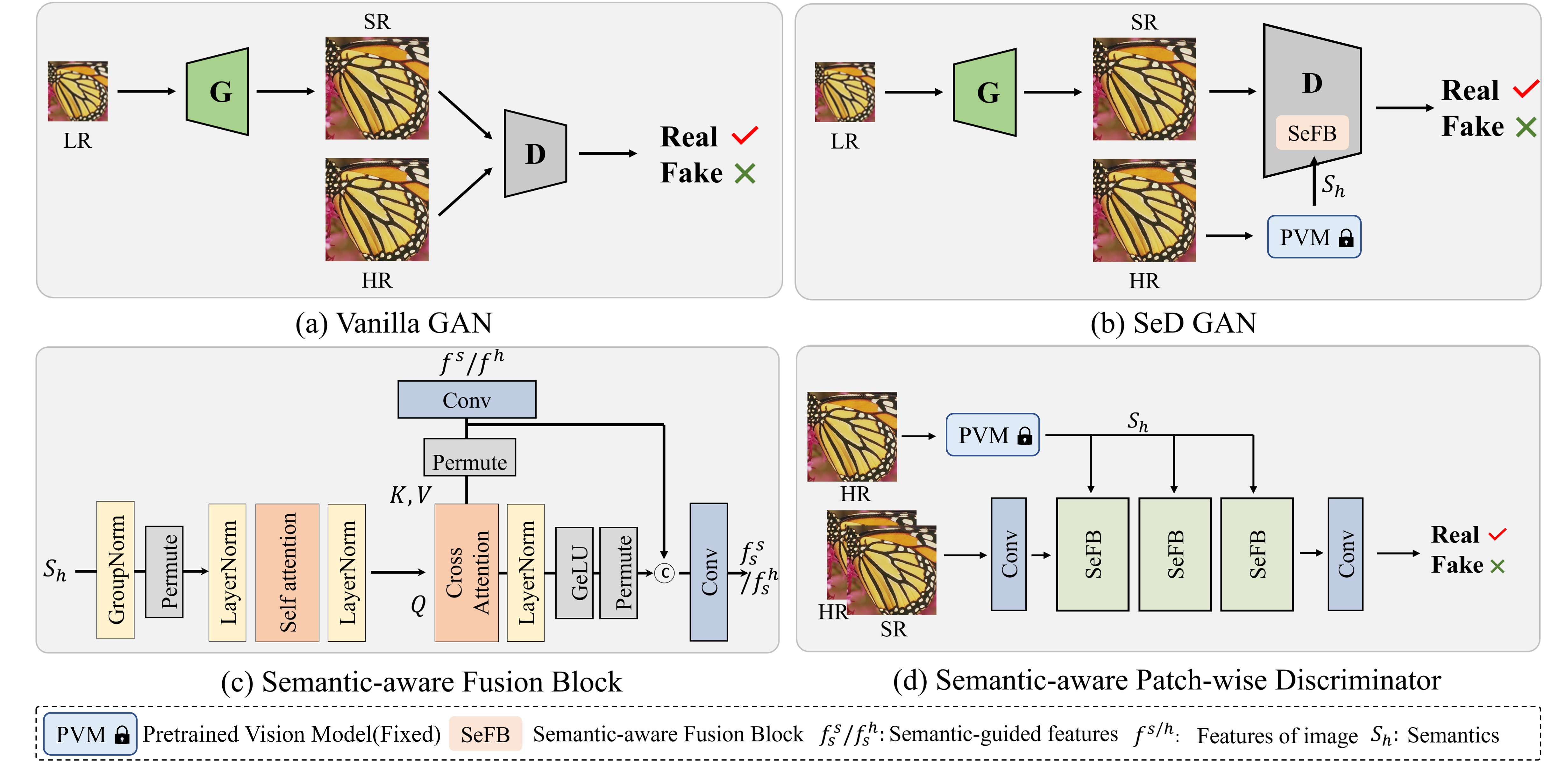}
    \caption{Illustration of (a) GAN-based SR with the vanilla discriminator. (b) Our proposed semantic-aware discriminator (SeD). (c) The network structure of SeFB. (d) The network structure of P+SeD. The vanilla discriminator measures the distributions of images regardless of the semantics, which causes the SR network to learn the average textures (\ie, noise) or generate textures not related to the semantics. In contrast, our proposed semantic-aware discriminator exploits the fine-grained semantics as the condition of the discriminator, which poses the SR network to learn more fine-grained semantic-aware textures for SR.}
    \label{fig:framework}
\end{figure*}

\section{Related Works}

\subsection{Single Image Super-resolution}
In recent years, deep learning has posed great progress in Single Image Super-resolution (SISR)~\cite{SRCNN,RDN}. Typically, most SR methods focus on exploring better backbones~\cite{EDSR,SAN,RCAN,swinir,HST,pang2020fan,aim2022} to improve objective quality. 

As the pioneering work,~\cite{SRCNN} firstly propose the SRCNN, which introduces the CNN to SISR and optimizes the network by minimizing the mean square error (MSE) between the super-resolved images (SR) and their corresponding high-resolution (HR) counterparts. Subsequently, numerous SR works have been explored to further improve the representation capability of the network from different perspectives, such as incorporating the residual connection~\cite{SRGAN,EDSR,VDSR}, dense connection~\cite{RDN,ESRGAN}, multi-scale representation~\cite{MSRN,MSDNN,}, etc. Specifically, some works~\cite{RCAN,SAN, SPARNet} employ attention mechanisms to further enhance high-frequency textures. However, these works lack the utilization of long-range dependencies present in images, thereby limiting their performance on SR. With the emergence of vision transformers, some studies~\cite{TTSR,swinir,ESRT} regard pixels as tokens and make full use of long-range dependencies with window-based self-attention, resulting in visually pleasing outputs for both classical (\ie, bicubic) and real-world image SR tasks.

\subsection{GAN-based Image Super-resolution}
To generate more appealing textures, some works aim to leverage the excellent hallucination capability of the generative models to enhance the subjective quality of SR Networks.  In contrast to commonly-used pixel-wise constraints, \eg, L1 and L2 losses, the adversarial loss is introduced to enable the SR network to learn the distribution of real-world high-quality images. SRGAN~\cite{SRGAN}, as the first work for GAN-based SR, systematically investigates the advantages of GAN for SR. Following this, a series of works~\cite{ranksrgan,ESRGAN,GAN-sisr1,LapGAN} began to refine the architecture of the SR network (\ie, the generator), achieving stable training and more realistic textures. Despite this, few works take into account optimizing another core component in GAN, \ie, the discriminator, which determines whether the distribution learned by the SR network aligns with real-world high-quality images. 

Starting from this point, our proposed Semantic-aware Discriminator (SeD) is expected to enable the SR network to achieve more fine-grained semantic-aware texture generation. 

\subsection{Pretrained Vision Models (PVMs)}
Benefiting from large-scale datasets, such as ImageNet-22K~\cite{imagenet}, JFT-300M~\cite{jft}, LAION-5B~\cite{LAION}, and the elaborate pretraining strategies~\cite{BERT,KDEP}, Pretrained Vision Models (PVMs), \eg, ViT~\cite{ViT,ViT2}, Swin Transformer~\cite{swin,swinV2}, have demonstrated their immense potential in various fields, \eg, classification~\cite{PALI,Coca}, vision-language task~\cite{CLIP,NLX-GPT,GPT-3,li2024graphadapter}, generation tasks~\cite{DaLLE,LDM,Taming}. Compared to smaller models~\cite{resnet}, PVMs are friendly to most down-streaming tasks, thanks to their strong and robust representation capability. Recently, the large vision-language model CLIP~\cite{CLIP} garnered significant interest for leveraging extensive amounts of collected vision-language pairs to learn a more fine-grained semantic space. As a result, a series of works incorporate this powerful feature extractor of CLIP into various tasks, including prompt learning~\cite{prompt}, generation, etc. Inspired by this, we aim to extract more fine-grained semantics from CLIP to guide our SeD.

\section{Method}
In this section, we first review the typical GAN-based SR method and then describe the difference between our method with it in the overall framework. Next, we clarify our proposed semantic-aware discriminator in detail. Finally, we describe how to integrate our SeD into the existing patch-wise discriminator and pixel-wise discriminator.

\subsection{Preliminary}
\label{sec:preliminary}
GAN-based SR \cite{SRGAN,ESRGAN,BSRGAN,RealESRGAN,LapGAN,cal-gan} aims to improve the perceptual quality of super-resolved images. With a simple discriminator, GAN-based SR introduces adversarial training to enable the SR network to generate vivid textures. The objectives of GAN-based SR are composed of three losses, including pixel-wise supervised loss $\mathcal{L}_s$, \eg, L1 Loss and MSE loss, a perceptual loss $\mathcal{L}_p$, and an adversarial loss $\mathcal{L}_{adv}$. Among them, pixel-wise supervised loss is used to constrain the pixel-wise consistency and ensure the realness of pixel values. Perceptual loss~\cite{Perceptualloss,SRGAN} $\mathcal{L}_p$ exploits the features of VGG~\cite{VGG} to give coarse perception constraints between generated and ground-truth images. In the adversarial training process, a discriminator $D$ is exploited to distinguish whether the image is real or fake, which is optimized with the loss function as:
\begin{equation}
    \mathcal{L}_D = \mathbb{E}_{I_h \sim P_{I_h}}log(1-D(I_h))+\mathbb{E}_{I_s \sim P_{I_s}}(D(I_s)),
\end{equation}
where the $P_{I_h}$ and $P_{I_s}$ are the distribution of the high-quality images $I_{I_h}$ and the super-resolved images $I_s$, respectively. The SR network (\ie, the generator) is optimized by the combination of three losses as:
\begin{equation}
    \mathcal{L}_G = \mathcal{L}_s + \lambda_p \mathcal{L}_p + \lambda_a \mathcal{L}_{adv},
\end{equation}
where adversarial loss $\mathcal{L}_{adv}$ enables the generator to cheat the discriminator $D$ with the contrary purpose of $\mathcal{L}_D$ as:
\begin{equation}
    \mathcal{L}_{adv} = \mathbb{E}_{I_h \sim P_{I_h}}log(D(I_h))+\mathbb{E}_{I_s \sim P_{I_s}}(1-D(I_s)).
\end{equation}

\subsection{Overall Framework}
\label{sec:overallframework}
The overall framework of our proposed Semantic-aware Discriminator (SeD) is shown in Fig.~\ref{fig:framework}. Given the low-resolution image $I_l$, we can first obtain the super-resolved image $I_s$. Then a discriminator $D$ is used to distinguish the $I_s$ and high-resolution image $I_h$, which enforces the SR network to generate the real-like images (\ie, $P(I_s)\approx P(I_h)$). However, the vanilla discriminator only takes into account the coarse-grained distribution of images, while ignoring the semantics of images. This will cause the SR network to produce fake and even worse textures. A promising texture generation should satisfy its semantic information. Therefore, we aim to achieve the semantic-aware discriminator, which leverages the semantics of high-resolution image $I_h$ as a condition. Here, we represent the semantic extractor in the large vision model as $\phi$, and we aim to achieve more fine-grained semantic-aware texture generation, which targets for $P(I_s|\phi(I_h))=P(I_h|\phi(I_h))$.

Therefore, as shown in the Fig.~\ref{fig:framework}, the high-resolution image $I_h$ will be fed into a fixed pretrained semantic extractor from large vision model $\phi$ to obtain the semantics $\phi(I_h)$. Then a semantic-aware fusion block (SeFB) is used to further warp the super-resolved image feature $f^s$ and high-resolution image feature $f^h$ to the discriminator as $f_s^s$, $f_s^h$ by regarding the semantics as the query. Based on the semantic-aware features, the discriminator can achieve the semantic-aware distribution measurement. Notably, this process does not increase the parameters or computational cost for the SR network in the inference stage, as discriminator is no longer needed. In the next section, we will clarify our semantic-ware discriminator in detail. 

\subsection{Semantic-aware Discriminator}
\label{sec:sed}
Our semantic-aware discriminator consists of two essential components, \ie, the semantic feature extractor and the semantic-aware fusion block (SeFB). By incorporating these components into the popular discriminators, we can obtain our Semantic-aware discriminators. 

\subsubsection{Semantic Excavation}
\label{sec:semantic_extractor}
A successful semantic excavation plays an irreplaceable role in the Semantic-aware Discriminator. The intuition is that more fine-grained semantics will promote the discriminator to measure finer-grained distributions. Therefore, the recently popular pretrained vision models (PVMs), which possess more powerful representation capabilities, are proper for semantic excavation. Among them, the CLIP model has emerged as a prominent backbone and has been applied in various tasks~\cite{DaLLE,CLIP-art,rombach2022text,lu2023beyond}. Consequently, we adopt the pretrained CLIP "RN50" model as the semantic extractor, since we require the pixel-wise semantic information for SR. Specifically, the ``RN50" is composed of four layers, with the resolution of features being down-sampled as the layers increase and the semantics becoming more abstract.
To investigate which layer is more suitable for our semantic excavation, we systematically conduct experiments for these four layers, and experimentally find that the semantic features from the third layer are optimal. Here, we represent the semantic extractor as $\phi$, and the semantic features $f_s$ are extracted by feeding the high-resolution image $I_h$ into $\phi$.

\subsubsection{Semantic-aware Fusion Block}
\label{sec:sefb}
After obtaining the semantics $S_h$, we exploit our proposed semantic-aware fusion block (SeFB) to implement the semantic guidance of Discriminator. The architecture of SeFB is shown in Fig.~\ref{fig:framework}(c). Given the features of the super-resolved/high-quality images $f^s$/$f^h$, we aim to warp the semantic-aware textures from images to the discriminator, which enforce the discriminator to focus on the distribution of semantic-aware textures. Therefore, in Fig.~\ref{fig:framework}(c), the semantics $S_h$ is passed to the self-attention module and then is fed to the cross-attention module as the query: 
\begin{equation}
    Q = \mathrm{LN}(\mathrm{SA}(\mathrm{LN}(\mathrm{GN}(S_h)))),
\end{equation}
where $\mathrm{LN}$, $\mathrm{GN}$, $\mathrm{SA}$ denotes the layer normalization, group normalization, and self-attention module. 
Then, the features of super-resolved/high-quality images $f^s/f^h$ are passed into the convolution layer and are regarded as the keys $K^s$/$K^h$ and values $Q^s$/$Q^h$. Finally, the warped semantic-aware image features with cross-attention are concatenated with original enhanced features to form the final features $f^s_s$/$f^h_s$ as:
        \begin{align}
        & {f^{(*)'}_s} = Softmax(Q^{(*)}K^{(*)T}/\sqrt{d_k})V^{(*)} \\ \notag
        & {f^{(*)}_s} = \mathrm{Conv}(\mathrm{Concat}(\mathrm{GELU}(\mathrm{LN}(f^{(*)'}_s)), \mathrm{Conv}(f^{(*)}))),
        \end{align}
    where ${f^{(*)}_s}$ denote the warped semantic-aware image features. $*$, $d_k$, and $\mathrm{GELU}$ are $s/h$,  scale factor, and  
 the Gaussian Error Linear Units, respectively.  $\mathrm{Conv}$ and $\mathrm{Concat}$ represent the convolution layer and the concatenate operation.

\subsection{Extension to Various Discriminators}
\label{sec:extension}
In this paper, we incorporate our proposed SeD into two popular discriminators, including a patch-wise discriminator and a pixel-wise discriminator. As illustrated in Fig.~\ref{fig:framework}(d), the patch-wise semantic-aware discriminator consists of three SeFBs and two convolution layers. For the pixel-wise discriminator, we follow the approach in~\cite{RealESRGAN} and exploit the U-Net architecture as the backbone. We substitute original convolution layers with our proposed SeFBs in the shallow feature extraction stage. 

\begin{figure*}[ht!]
    \centering
    \includegraphics[width=1.\linewidth]{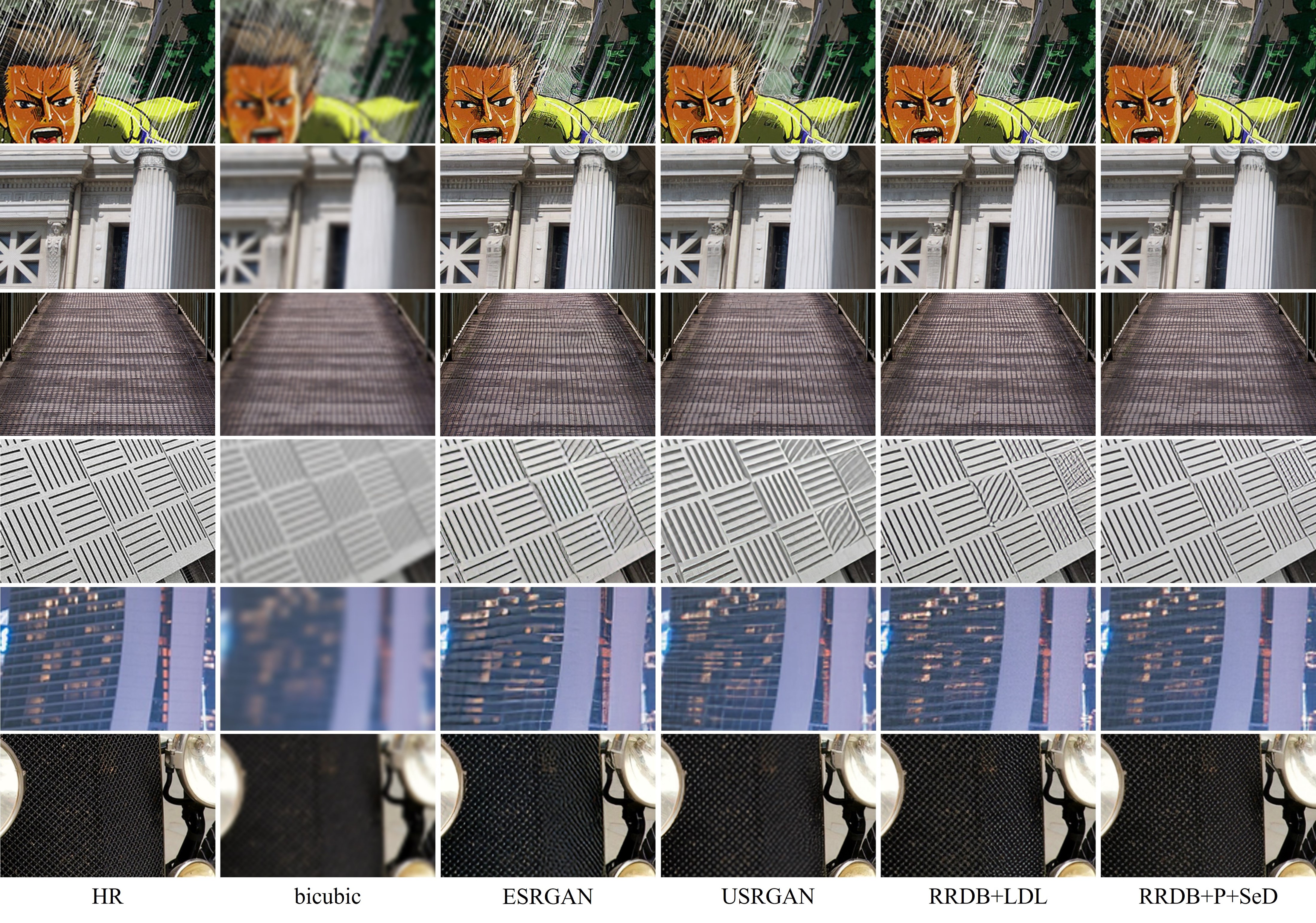}
    \caption{Visual comparison (zoom-in for better view) to state-of-the-art GAN-based SR methods. We demonstrate patch-wise SeD here because it shows better subjective quality. With SeD, the SR network is capable of restoring photo-realistic textures.}
    \label{fig:qualitative_classic}
\end{figure*}

\section{Experiments}
\subsection{Experiment setups}
\noindent \textbf{Network structures.} 
For the generator, we use the competitive RRDB~\cite{ESRGAN} as the backbone. To demonstrate the generalization ability of our proposed SeD, we also employ the popular transformer-based SR network SwinIR~\cite{swinir} as our generator. For the discriminator, we incorporate our SeD into two types of discriminators, \ie, the patch-wise discriminator~\cite{patchgan} and the pixel-wise discriminator~\cite{unet-d}, namely, P+SeD and U+SeD. We also validate our SeD on an image-wise discriminator based on VGG~\cite{ESRGAN}, dubbed V+SeD, which is provided in the \textbf{Appendix}.

\begin{table*}[h!]
\begin{center}
\resizebox{\textwidth}{!}{
\begin{tabular}{l|ccccc}
\toprule
\textbf{Method}  & Set5 & Set14 & DIV2K & Urban100 & Manga109\\ 
\hline
SFTGAN  & 0.080/30.06/0.848  &  -/-/-   &  0.133/28.08/0.771   & 0.134/24.34/0.723 & 0.072/28.17/0.856  \\
ESRGAN  & 0.076/30.44/0.852  &  0.133/26.28/0.699   &  0.115/28.20/0.777   & 0.123/24.37/0.734 & 0.065/28.41/0.859  \\
USRGAN  & 0.079/30.91/0.866  &  0.143/27.15/0.736   &  0.132/28.77/0.793   & 0.133/24.89/0.750 & 0.063/28.75/0.872  \\
RRDB+LDL  & 0.069/31.03/0.861  &  0.121/26.94/0.721   &  0.101/28.95/0.795   & 0.110/25.50/0.767 & 0.055/29.41/0.875  \\
RRDB+DualFormer  & 0.068/31.40/0.872  &  0.121/27.53/0.741   &  0.103/29.30/0.802   & 0.115/25.73/0.774 & 0.053/29.90/0.886  \\
\hline
RRDB+P  & 0.070/30.67/0.860  & 0.130/26.92/0.724    &  0.111/28.71/0.792   & 0.120/24.86/0.752 & 0.058/28.60/0.872  \\
 RRDB+P+SeD  & \textbf{0.064}/31.22/0.867  & \textbf{0.117}/27.37/0.736    & \textbf{0.094}/29.27/0.802   & \textbf{0.106}/25.93/0.779 & \textbf{0.048}/29.99/0.888  \\
\hline
RRDB+U & 0.072/31.13/0.869 & 0.127/27.52/0.739 & 0.110/29.28/0.802 & 0.125/25.61/0.768 & 0.056/29.49/0.882 \\
 RRDB+U+SeD  & \textbf{0.069}/31.73/0.880  & \textbf{0.123}/27.94/0.757    &  \textbf{0.102}/29.85/0.818  & \textbf{0.112}/26.20/0.788 & \textbf{0.047}/30.46/0.897 \\
\hline\hline
SwinIR+LDL  & 0.065/31.03/0.861  &  0.118/27.22/0.732  &  0.094/29.12/0.801   & 0.102/26.23/0.792 &  0.047/30.14/0.888 \\
\hline
SwinIR+P  & 0.070/31.49/0.876  &  0.127/27.66/0.747  &  0.103/29.66/0.815   & 0.107/26.22/0.790 &  0.048/30.18/0.895 \\
 SwinIR+P+SeD  & \textbf{0.061}/31.44/0.870  &  \textbf{0.115}/27.53/0.742   &  \textbf{0.090}/29.53/0.810   & \textbf{0.097}/26.45/0.794 & \textbf{0.044}/30.48/0.896 \\
\hline
SwinIR+U  & \textbf{0.064}/31.38/0.869  &  0.120/27.64/0.744   &  \textbf{0.095}/29.56/0.810   &  0.103/26.09/0.786 &  0.049/29.99/0.889 \\
 SwinIR+U+SeD  & 0.067/31.64/0.874  &  \textbf{0.117}/27.84/0.750   &  0.096/29.79/0.816 &  \textbf{0.102}/26.46/0.796 &  \textbf{0.045}/30.58/0.898 \\
\bottomrule
\end{tabular}}
\end{center}
\caption{Quantitative comparison between GAN-based SR methods and the proposed SeD. Here, we use ``+P" to denote a PatchGAN discriminator; ``+U" to denote a U-Net discriminator; ``+SeD" to denote our implementation based on two different discriminator architectures.
The best perceptual results of each group are highlighted in bold. Each result is in term of LPIPS$\downarrow$/PSNR$\uparrow$/SSIM$\uparrow$, $\uparrow$ and $\downarrow$ mean that the larger or smaller score is better, respectively.}
\label{table:classical}
\end{table*}

\noindent \textbf{Implementation details for classical image SR.} We conduct our experiments with SeD on two typical SR tasks, including classical image SR~\cite{timofte2017ntire} and real-world image SR~\cite{RealESRGAN}. The classical image SR aims to super-resolve the low-resolution images with the bicubic downsampling. Following previous works~\cite{DBPN,swinir}, we train our network using 3450 images from DF2K~\cite{DIV2K,Flickr2K}. We test our methods on five benchmarks: Set5~\cite{Set5}, Set14~\cite{Set14}, Urban100~\cite{urban100}, Manga109~\cite{manga109} and DIV2K validation set~\cite{DIV2K}. We synthesize all the $\times 4$ training samples using MATLAB bicubic downsampling kernel. We apply the same data augmentation to all the training samples following previous arts~\cite{li2020learning,swinir,aim2022}. The patch size of HR images is 256 $\times$ 256. We implement the experiments on 4 NVIDIA V100 GPUs with PyTorch and a batch size of 8 on each device. The parameters of the generator are initialized with pretrained PSNR-oriented model. We use Adam optimizer with a learning rate of 1e-4 to optimize our network. The total training iterations are 300k.

\noindent \textbf{Implementation details for real-world image SR.} We also perform experiments on real-world image SR. We compare the performance with the original discriminator (\ie, without semantics) on three state-of-the-art methods: Real-ESRGAN~\cite{RealESRGAN}, LDL~\cite{LDL} and SwinIR~\cite{swinir}. Following them, we evaluate the performance on several commonly-used  real-world low-resolution datasets, including DPED~\cite{DPED}, OST300~\cite{SFTGAN} and RealSRSet~\cite{BSRGAN}. The training strategy and dataset of the three methods are slightly different. To keep fairness, we follow their original training settings to train the generator with our SeD, and compare the visual quality with their original GAN-based results. Furthermore, we adopt the well-known no-reference metric NIQE~\cite{NIQE} for the quantitative comparison, since the ground-truth images are not available in the real world.
\begin{figure*}[ht!]
    \centering
    \includegraphics[width=1.\linewidth]{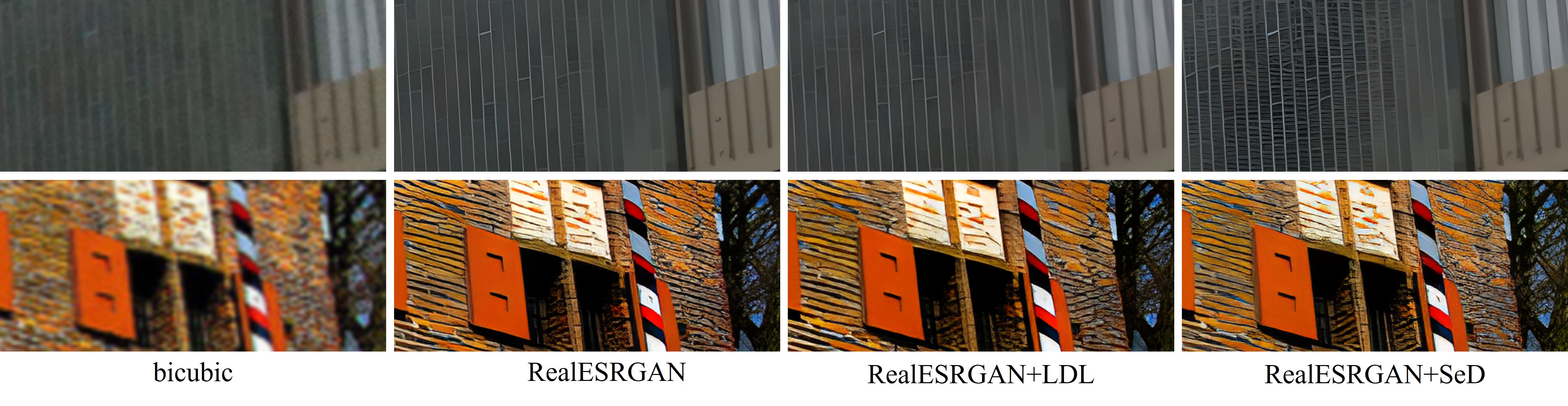}
    \caption{Visual comparison (zoom-in for better view) to state-of-the-art GAN-based real-world SR methods. We demonstrate pixel-wise SeD here to align with previous works~\cite{RealESRGAN,LDL}.}
    \label{fig:qualitative_real}
\end{figure*}

\subsection{Results on classical image SR}
\label{sec:csr}
We conduct a quantitative comparison between our SeD and the state-of-the-art GAN-based SR methods. SFTGAN~\cite{SFTGAN} incorporates a semantic map extracted from input LR images to better restore textures. ESRGAN~\cite{ESRGAN} and USRGAN~\cite{USRNet} both use RRDB~\cite{ESRGAN} as the backbone of the generator. LDL~\cite{LDL} employs an artifact map into discriminator with RRDB to better reconstruct finer textures. Therefore, we exploit the RRDB as the generator for a fair comparison with them. To validate the generalization capability, 
we also validate our SeD with another generator architecture, \ie, transformer-based SR backbone, SwinIR~\cite{swinir}.  

As demonstrated in Table~\ref{table:classical}, our SeD outperforms existing GAN-based methods in the perception metric (\ie, LPIPS), including SFTGAN~\cite{SFTGAN}, ESRGAN~\cite{ESRGAN}, USRGAN~\cite{USRNet}, LDL~\cite{LDL} and DualFormer~\cite{Dualformer}, with the comparable and even more high objective qualities. It is noteworthy that, for the last several rows of the Table~\ref{table:classical}, we validate the effectiveness of our SeD by incorporating our SeD into different generators (\ie, RRDB~\cite{ESRGAN} and SwinIR~\cite{swinir}) and discriminators (patch-wise discriminator, \ie, ``+P" and U-shape pixel-wise discriminator, \ie, ``+U"). For ``RRDB+P'', we can observe that our SeD (\ie, ``RRDB+P+SeD") outperforms the vanilla discriminator in LPIPS under all circumstances, even with up to 1.0dB PSNR gains in Manga109 dataset. This indicates the superiority of incorporating semantic guidance in our SeD, as it can discriminate more fine-grained textures during training, resulting in better perceptual quality. Notice that, our SeD also performs well on the transformer-based SR method SwinIR (\ie, ``SwinIR+P+SeD", and ``SwinIR+U+SeD"). This demonstrates the generalization capacity of SeD on different generator architectures.

Fig.~\ref{fig:qualitative_classic} presents the visual comparison between state-of-the-art methods and our SeD on classical image SR. We can notice that other methods restore images with virtual textures, \eg, twisted rain drops in the first row; unnatural textures on the surface of the bridge in the second row; mistaking the dotted texture in front of the car for stripes in the last row, \etc. Compared with other methods, our SeD generates visual-pleasant results without introducing artifacts. This proves that incorporating semantic guidance through SeD can empower the discriminator to distinguish finer details. We provide more visual results in the \textbf{Appendix}.

\subsection{Results on real-world image SR}
\label{rsr}
We evaluate the performance of SeD on the real-world image SR problem by incorporating it into state-of-the-art methods, including Real-ESRGAN~\cite{RealESRGAN} and LDL~\cite{LDL}. Typically, Real-ESRGAN and LDL both utilize RRDB as a generator, but LDL replaces the vanilla U-Net discriminator with there artifact mapped one. We also conduct experiments with a more powerful transformer-based generator, SwinIR~\cite{swinir}, the qualitative results of SwinIR are provided in the \textbf{Appendix}. All of the above-mentioned methods utilize the pixel-wise U-Net discriminator, therefore, we conduct our experiments with the pixel-wise discriminator ``U+SeD".

We demonstrate the quantitative comparison in Table~\ref{table:real}. Our SeD surpasses the vanilla discriminator on most of the real-world datasets in terms of different backbones and training strategies. The results suggest that our SeD is capable of discriminating heavily distorted real-world images, and learning the fine-grained texture generation ability. As shown in Fig.~\ref{fig:qualitative_real}, for the first line, images restored by SeD contain more realistic textures; for the second line, SeD eliminates the smooth problem and generates vivid textures for bricks.
We provide more  visualizations in the \textbf{Appendix}.

\subsection{Ablation studies for SeD}
We conduct ablation studies to investigate the most suitable way of introducing semantic guidance into discriminator. We focus on three aspects: i) What is the best way to fuse semantic features in SeFB? ii) Which layer of CLIP contains the most useful semantic information for texture discrimination? iii) Which pretrained model is more suitable as a semantic extractor?

\begin{table}
\begin{center}
\setlength{\tabcolsep}{0.3mm}{\begin{tabular}{l|c|c|c|c}
\toprule
\textbf{Method} &  \textbf{Type} & DPED & OST300 & RealSRSet \\

\hline
 \multirow{3}{*}{RealESRGAN} & V & 5.27 & 2.82 & 5.80 \\
  & L & 5.36 & 2.83 & 6.07 \\
 & S & \textbf{4.49} & \textbf{2.73} & \textbf{5.34} \\
 \hline
 \multirow{3}{*}{SwinIR} & V & \textbf{4.95} & 2.93 & 5.49 \\
 & L & 5.65 & 3.10 & 5.59 \\
 & S & 5.16 & \textbf{2.75} & \textbf{5.15} \\
\bottomrule
\end{tabular}}
\end{center}
\caption{The quantitative comparisons of our proposed SeD with the state-of-the-art methods on Real-world SR in terms of NIQE$\downarrow$. ``V" denotes ``vanilla discriminator", ``S" denotes ``SeD", and ``L" denotes ``LDL".}
\label{table:real}
\end{table}

\begin{table}[h]
\begin{center}
\begin{tabular}{l|cccc}
\toprule
\multirow{2}{*}{Datasets} &  \multicolumn{4}{c}{Fusion methods (LPIPS$\downarrow$)} \\ \cline{2-5}  & SeD-A & SeD-B & SeD-C & SeD-our\\ 
\midrule
Set5 &  0.069  &  0.085 & 0.074   &   \textbf{0.064}   \\
Urban100  &  0.124 &  0.131   &   0.125  & \textbf{0.106}   \\
Manga109 &  0.058 &  0.065   &  0.055   &   \textbf{0.048} \\
\bottomrule
\end{tabular}
\end{center}
\caption{Ablation studies for different fusion methods in SeFB. We perform all the experiments on classical image SR, and evaluate with subjective metric LPIPS. ``A'', ``B'', ``C'' denote for concatenate, channel attention, spatial attention, respectively.}
\label{ab:fusion}
\end{table}
\begin{table}
\begin{center}

\begin{tabular}{l|cccc}
\toprule
\multirow{2}{*}{Methods}    & \multicolumn{3}{c}{Datasets (LPIPS$\downarrow$)} \\ 
\cline{2-4}  &  Set5 &  Urban100 &Manga109\\ 
\midrule

SeD-Layer1  & 0.066  &  0.120  &  0.057     \\
SeD-Layer2    & 0.068  & 0.117  & 0.056   \\
SeD-Layer3    &  \textbf{0.064}  &   \textbf{0.106}  &   \textbf{0.048} \\
SeD-Layer4   &   0.069  &  0.122   &  0.055   \\
\bottomrule
\end{tabular}
\end{center}
\caption{Ablation for the semantics from different layers of CLIP.}
\label{table:layers}
\end{table}

\begin{table}
\begin{center}

\begin{tabular}{l|ccc}
\toprule
\multirow{2}{*}{Method}    & \multicolumn{3}{c}{Datasets (LPIPS$\downarrow$)} \\ 
\cline{2-4}  &  Set5 &  Urban100 & Manga109\\ 
\midrule
ResNet-50  & 0.066  &  0.117  &  0.055     \\
CLIP    &  \textbf{0.064}  &   \textbf{0.106}  &   \textbf{0.048} \\
\bottomrule
\end{tabular}
\end{center}
\caption{Ablation studies for different semantic extractors. }
\label{table:extractors}
\end{table}

\noindent\textbf{Effects of different fusion methods.} There are a series of fusion methods to introduce the priors, including feature concatenation, spatial attention, or channel attention. We compare our proposed SeFB (\ie, SeD-Our) with different fusion strategies in Fig.~\ref{ab:fusion}. The channel attention (\ie, SeD-B) destroys the spatial information of semantics, which achieves the most worse performance. The performance of concatenation operation (\ie, SeD-A) and spatial attention (\ie, SeD-C) are comparable, which is obviously lower than our proposed semantic-aware fusion block (SeFB). More details can be found in the \textbf{Appendix} for the implementations of the above fusion strategies. 

\noindent\textbf{The semantics of different layers.}
It is noteworthy that the semantics from different layers of CLIP~\cite{CLIP} is different. With the increase of layers, the semantics will be more representative while the resolution becomes lower. The essence of effective semantic guidance lies in extracting semantic features that are sufficiently deep yet retain essential spatial information to ensure the quality of restoration guidance. There are in total four layers inside CLIP semantic extractor, and the spatial resolution is halved after each layer. To investigate which layer is the best choice for the guidance, we conduct the experiments in Table ~\ref{table:layers}. From the table, In the first and second layers, despite the semantic feature possessing a relatively large spatial resolution, its lack of concentration could result in suboptimal semantic guidance for the discrimination process. For the last layer, the spatial resolution is too small, which may causes discriminator ignoring the pixel-wise consistency. As the result, the semantics from the third layer performs best, which is used in our paper. 

\noindent\textbf{The effects of the different semantic extractors.}
To delve deeper into the question of whether finer-grained semantic understanding enhances guidance quality, we compare two distinct semantic extractors. Specifically, we evaluate a pretrained ResNet-50~\cite{resnet} using the ImageNet~\cite{imagenet} dataset against the "RN50" model from the pretrained CLIP, which utilizes a vision-language dataset, as detailed in Table.~\ref{table:extractors}. The results of our experiments indicate that the "RN50" model pretrained with vision-language data outperforms the former. This superior performance is attributed to the language descriptions in CLIP's training regimen, which provide more comprehended and detailed semantic context compared to traditional digital labels used in classification tasks. This finding underscores the potential of language-enhanced models in achieving a more refined understanding of semantics, thereby offering more effective guidance in discriminative processes.

\subsection{T-SNE visualization of discriminator
features}
\begin{figure}[htbp]
    \centering
    \includegraphics[width=1.0\linewidth]{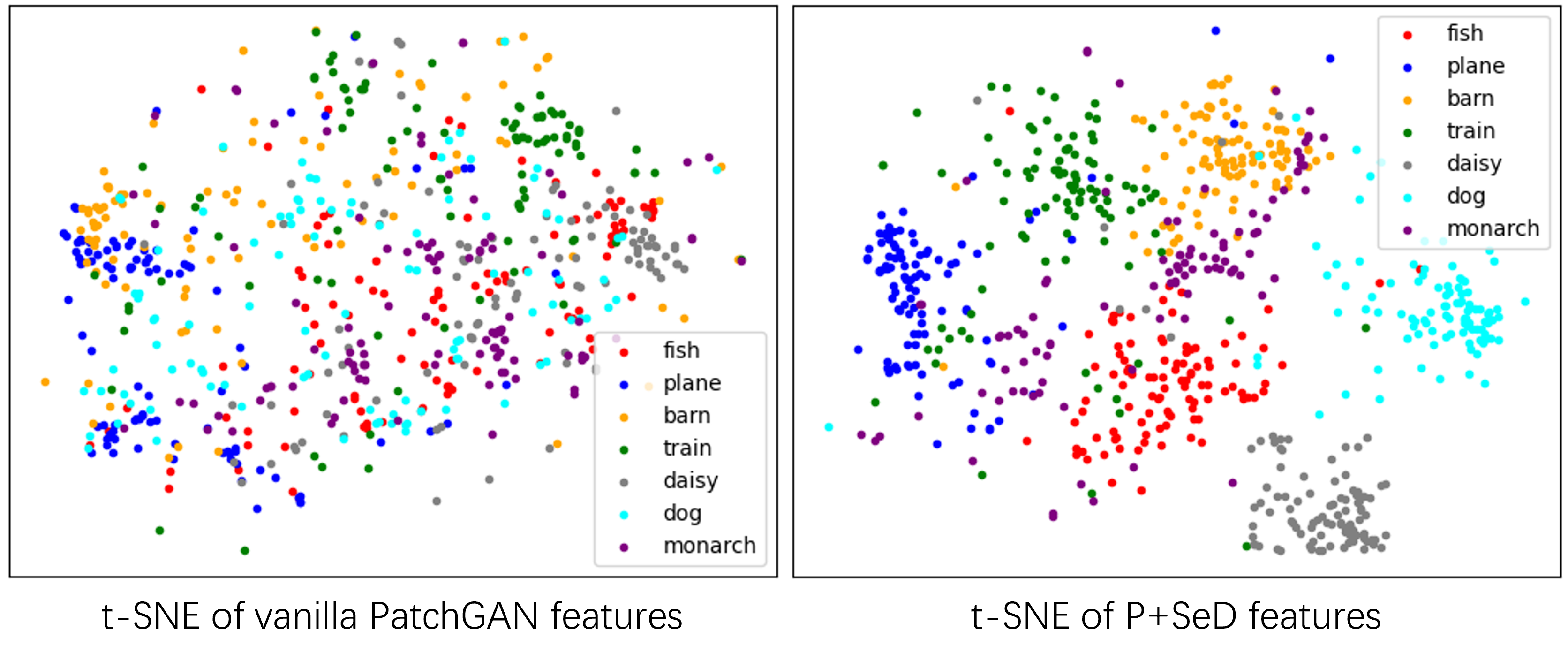}
    \caption{The t-SNE visualization of discriminator features. The 7 categories are ``fish", ``plane", ``barn", ``train", ``daisy", ``dog", ``monarch" from ImageNet, respectively.}
    \label{fig:t-sne}
\end{figure}
To validate that our semantic guidance promotes the discriminator to perform more fine-grained discrimination, we visualize the intermediate features for both vanilla discriminator and our SeD with t-SNE~\cite{t-sne}. We randomly select 7 categories from ImageNet~\cite{imagenet}, with 100 images each category. Then, we feed all the images into a vanilla PatchGAN discriminator and our P+SeD. We visualize the features after the first SeFB of SeD, and the features after the first BN layer of PatchGAN discriminator, respectively. As shown in Fig.~\ref{fig:t-sne}, the features of SeD are well-clustered, while the features of vanilla patch discriminator are disorganized. This provides the evidence that under the guidance of semantics, SeD is capable of performing more fine-grained discrimination, as claimed in the paper.

\section{Conclusion}
In this paper, we propose a simple but effective Semantic-aware Discriminator (SeD) for Image Super-resolution. We find the distribution learning of previous discriminators is overly coarse-grained, which causes the virtual or counter-intuitive textures. To mitigate this, we introduce the semantics of the image from pretrained vision models (PVMs) as the condition of the discriminator with our proposed semantic-aware fusion block (SeFB), which encourages the SR network to learn the fine-grained semantic-aware textures. Moreover, our SeD is easy to be incorporated into most GAN-based SR methods and achieves excellent performance.
Extensive experiments on classic SR and Real-world SR have demonstrated the effectiveness of our SeD.

{
    \small
    \bibliographystyle{ieeenat_fullname}
    \bibliography{main}
}

\clearpage
\section*{Appendix}

\noindent Section~\ref{sec:recent} includes two recently proposed large-scale high-quality benchmarks here for classical image SR evaluation: LSDIR~\cite{li2023lsdir} and HQ-50K~\cite{yang2023hq}.

\noindent Section~\ref{sec:image-wise} provides the network structures of pixel-wise SeD (U+SeD), image-wise SeD (V+SeD) and CLIP semantic extractor. Additionally, we demonstrate the quantitative comparison between ESRGAN~\cite{ESRGAN} and our implemented ``RRDB+V+SeD". 

\noindent Section~\ref{sec:fusion} demonstrates the details of different fusion methods of SeD. 

\noindent Section~\ref{sec:generator} presents the ablation studies of introducing semantic guidance into the generator.

\noindent Section~\ref{sec:visual} visualizes more results of classical image SR and real-world image SR.

\begin{figure*}[!htbp]
    \centering
    \includegraphics[width=1.0\linewidth]{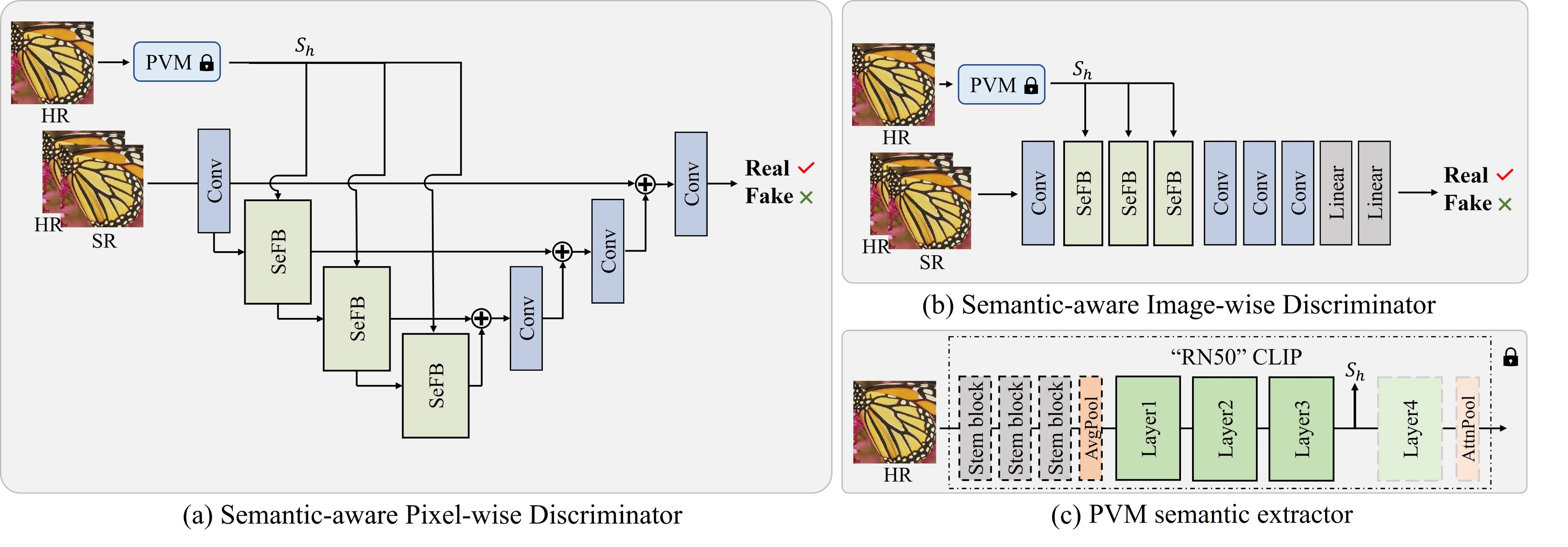}
    \caption{The framework of (a) pixel-wise U+SeD, (b) image-wise V+SeD, (c) ``RN50" CLIP semantic extractor.}
    \label{fig:v+sed}
\end{figure*}


\section{Evaluation on large-scale benchmarks}
\label{sec:recent}
With the development of image restoration, researchers have begun showing interest in larger-scale training and testing datasets, in addition to deeper model design. This interest is to align the field with other visual tasks, such as image recognition and image detection. To demonstrate the effectiveness of our SeD, we evaluate our methods on two recently proposed large-scale benchmarks, LSDIR~\cite{li2023lsdir} and HQ-50K~\cite{yang2023hq}. There are 250 available images from LSDIR and 1250 images from HQ-50K. Different from commonly used benchmarks (\eg, Set5~\cite{Set5}, Set14~\cite{Set14}, Urban100~\cite{urban100}, \etc), these testing images encompass a wide range of natural scenes, along with high resolution and complex textures. The results are shown in Table.~\ref{table:more_classical}. Notice that, we \textit{do not} train our model on training datasets of these two benchmarks. Instead, we directly use weights introduced in the main paper.

As demonstrated in the table, SeD outperforms in both objective and subjective metrics, indicating that semantic guidance is capable of not only better reconstructing simple textures in commonly used benchmarks~\cite{Set5,Set14,urban100,BSDall,manga109}, but also effectively handling complex textures in large-scale evaluation datasets. Qualitative comparisons are given in Sec.~\ref{sec:visual}.

\begin{table}[htbp]
\centering
\caption{Evaluation results of $\times$ 4 classical SR on large-scale benchmarks. Metrics are LPIPS$\downarrow$/PSNR$\uparrow$/SSIM$\uparrow$.}
\resizebox{\linewidth}{!}{
\begin{tabular}{l|cc}
\hline
Datasets & LSDIR~\cite{li2023lsdir} & HQ-50K~\cite{yang2023hq} \\ \hline
ESRGAN~\cite{ESRGAN} & 0.138/23.88/0.686 & 0.176/23.67/0.677 \\
RRDB+LDL~\cite{LDL} & 0.118/24.66/0.712 & 0.171/24.33/0.701 \\
RRDB+SeD & \textbf{0.116/25.20/0.727} & \textbf{0.157/24.66/0.710} \\ \hline
\end{tabular}}
\label{table:more_classical}
\end{table}

\section{Implementation detials of image-wise SeD}
\label{sec:image-wise}
We incorporate our proposed semantic-aware discriminator to a VGG-like discriminator~\cite{VGG}, dubbed V+SeD, which is shown in Fig.~\ref{fig:v+sed} (b).

In particular, the Image-wise discriminator has been explored in a series of GAN-based image SR networks~\cite{SRGAN,ESRGAN}, since it is simple and effective.

The quantitative results are demonstrated in Table~\ref{table:v+sed}. Our V+SeD significantly outperforms the vanilla VGG-like discriminator (which is used by ESRGAN~\cite{ESRGAN}) on both objective and subjective metrics. These further demonstrate the effectiveness and generalization capability of our suggested SeD with respect to various discriminator backbones.

\begin{table}[h!]
\begin{center}
{\begin{tabular}{l|cc}
\toprule
Datasets & ESRGAN & RRDB+V+SeD\\ 
\midrule
Set5  & 0.076/30.44/0.852  &  \textbf{0.070}/30.83/0.862   \\
Set14  & 0.133/26.28/0.699  &  \textbf{0.125}/27.06/0.729   \\
DIV2K  & 0.115/28.20/0.777  &  \textbf{0.107}/28.83/0.794   \\
Urban100 & 0.123/24.37/0.734  &  \textbf{0.118}/25.32/0.766   \\
Manga109  & 0.065/28.41/0.859  &  \textbf{0.057}/29.31/0.878   \\
\bottomrule
\end{tabular}}
\end{center}
\caption{Quantitative comparison between ESRGAN and V+SeD. The best perceptual results of each group are highlighted in bold. Each result is presented in terms of LPIPS$\downarrow$/PSNR$\uparrow$/SSIM$\uparrow$, $\uparrow$ and $\downarrow$ indicate that a larger or smaller score is better, respectively.}
\label{table:v+sed}
\end{table}

\section{Implementation details of different fusion strategies}
\label{sec:fusion}

\begin{figure}[!ht]
    \centering
    \includegraphics[width=1.0\linewidth]{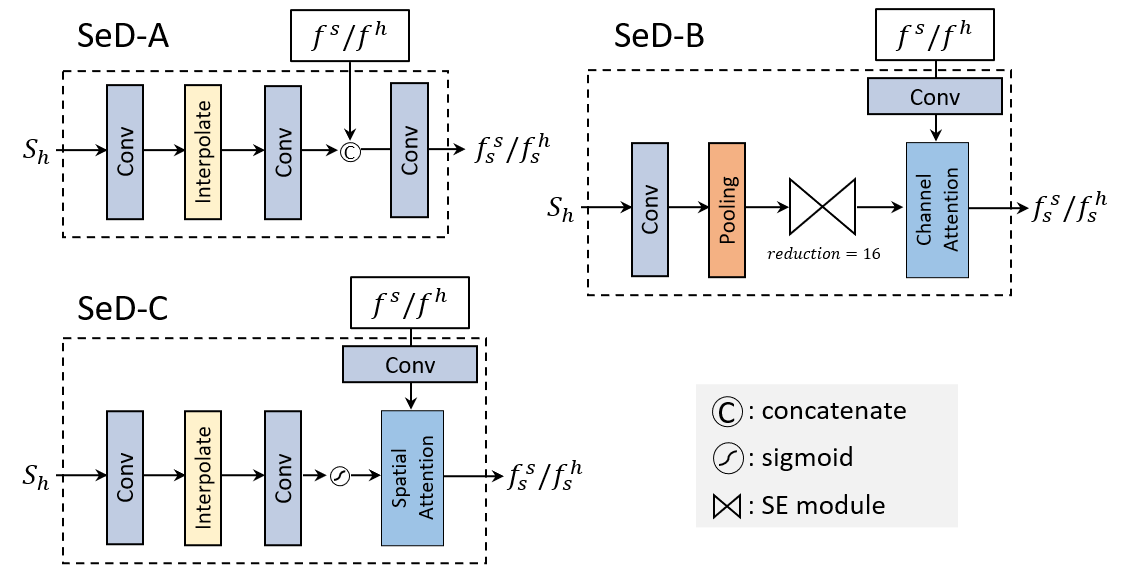}
    \caption{Frameworks of SeD-A, SeD-B, SeD-C, respectively.}
    \label{fig:SeFBs}
\end{figure}

We present the network architectures of our used SeD-A, SeD-B, and SeD-C in our ablation studies in Fig.~\ref{fig:SeFBs}. 
Among them, \textbf{SeD-A} uses concatenation operation as the fusion method.  \textbf{SeD-B} utilizes a channel-wise attention mechanism to fuse semantic information adaptively.  \textbf{SeD-C} leverages spatial-wise attention.  In contrast, our proposed SeD performs cross-attention between semantic feature and image feature, taking full advantage of abundant semantic information contained in LVMs, and maintaining the spatial information.  

\section{Ablation studies with Semantic-aware Generator}
\label{sec:generator}
As described in the main paper, one intuitive method to generate semantic-aware textures is to integrate the semantic guidance of images into the generator. To verify this idea, we conduct experiments of the semantic-aware generator on $\times$ 4 real-world image SR task. We choose RRDB~\cite{ESRGAN} with 11 residual in residual blocks as the baseline generator. For semantic-aware RRDB, namely Se-RRDB, we replace the $5^{th}$ and the $11^{th}$ block with the same semantic-aware fusion block we used in the SeD. We synthesize the degraded images by Real-ESRGAN~\cite{RealESRGAN} model, and evaluate the perceptual qualities of restored images on RealSR~\cite{RealSR} dataset in terms of NIQE~\cite{NIQE}. We first train two separate generators without discriminators, then fine-tune them on vanilla discriminator and our SeD, respectively. The results are shown in Table.~\ref{table:generator}. 

\begin{table}[h!]
\begin{center}
\setlength{\tabcolsep}{0.5mm}
{\begin{tabular}{c|cccc|cc}
\toprule
\# & RRDB &  Se-RRDB & Vanilla D & SeD & Canon & Nikon\\ 
\midrule
1 & \checkmark & & & & 7.56 & 7.83\\
2 & & \checkmark & & & 8.11 & 8.27\\
\hline\hline
3 &\checkmark & & \checkmark & & 4.71 & 5.14\\
4 & & \checkmark& \checkmark & & 5.39 & 5.66\\
\hline\hline
5 &\checkmark & & & \checkmark & 4.51 & 4.96\\
6 & & \checkmark&  & \checkmark& 4.78 & 5.32\\
\bottomrule
\end{tabular}}
\end{center}
\caption{Quantitative comparison between RRDB and Semantic-aware RRDB. $\checkmark$ means we use this backbone during training. The lower score is better.}
\label{table:generator}
\end{table}

As we can see, Se-RRDB performs worse than RRDB across all training paradigms, which reveals that incorporating the semantic information in the generator may not be appropriate for real-world image SR problems. The reason we guess is that it is difficult to extract accurate semantic information from low-quality images since the severe distortions in real world will cause the failure of the semantic extractor. Moreover, introducing the semantics into the generator will cause the catastrophic growth of computation complexity in the inference stage, where semantic extraction is costly and time-consuming.

Therefore, in our paper, we explore the more simple and effective semantic-aware discriminator (SeD), which improves the perceptual qualities of restored images, \ie,  the comparison between $3^{rd}$ and the $4^{th}$ lines or the comparison between the $5^{th}$ and the $6^{th}$ lines in Table.~\ref{table:generator}. Moreover, our SeD enables the SR network to restore more photo-realistic textures and does not require any additional computational burden during the inference stage.

\section{More visualization results}
\label{sec:visual}

\begin{figure}[h!]
\centering
\begin{subfigure}{0.23\linewidth}
\includegraphics[width=\linewidth]{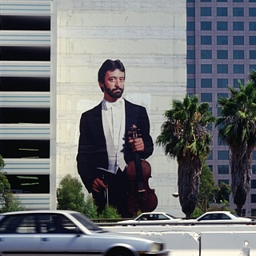}
\end{subfigure}
\hfill
\begin{subfigure}{0.23\linewidth}
\includegraphics[width=\linewidth]{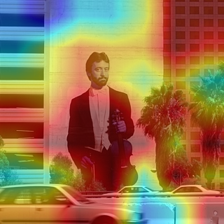}
\end{subfigure}
\hfill
\begin{subfigure}{0.23\linewidth}
\includegraphics[width=\linewidth]{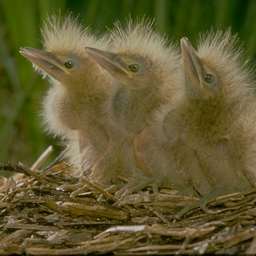}
\end{subfigure}
\hfill
\begin{subfigure}{0.23\linewidth}
\includegraphics[width=\linewidth]{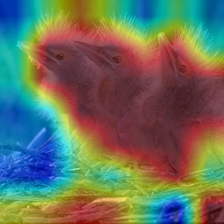}
\end{subfigure}
\caption{Visualization of semantics extracted from PVM.}
\label{fig:more_tsne}
\end{figure}

First, we show the visualization of semantics obtained from PVM in the image as Fig.~\ref{fig:more_tsne} (brighter region means more semantically important for PVM.) As stated, PVM can bring more fine-grained semantics in one image for guidance. (\eg, covering most semantics from trees, person, \etc. in the first sample). 

\begin{figure*}
    \centering
    \includegraphics[width=1.0\linewidth]{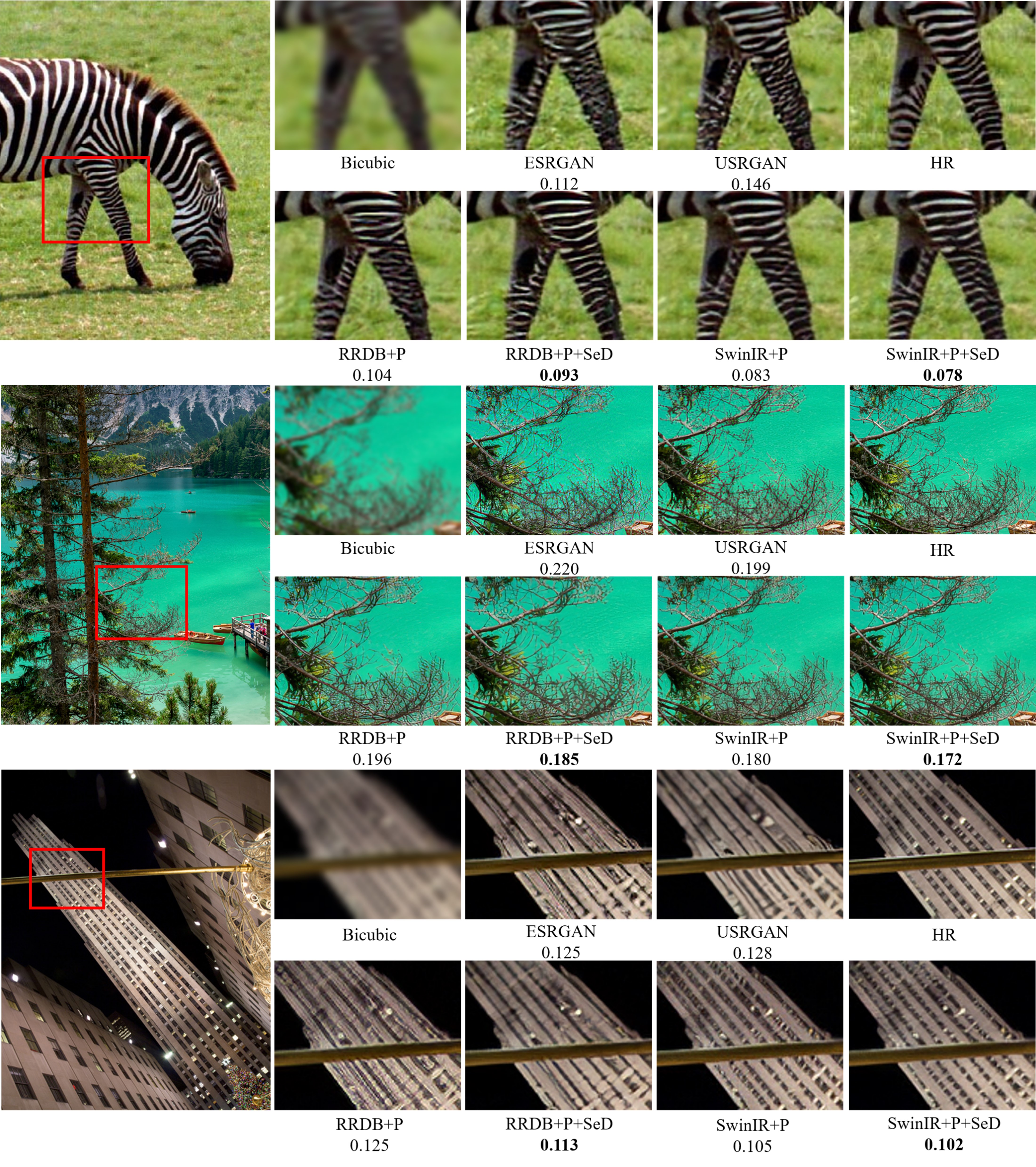}
    \caption{Visual comparisons between SeD and vanilla discriminator on classical image SR. To provide a clearer understanding of the perceptual qualities of images, we present the LPIPS$\downarrow$ here. It is evident that our SeD further enhances the ability of vanilla GAN to restore more realistic textures.}
    \label{fig:classical2}
\end{figure*}

\begin{figure*}
    \centering
    \includegraphics[width=1.0\linewidth]{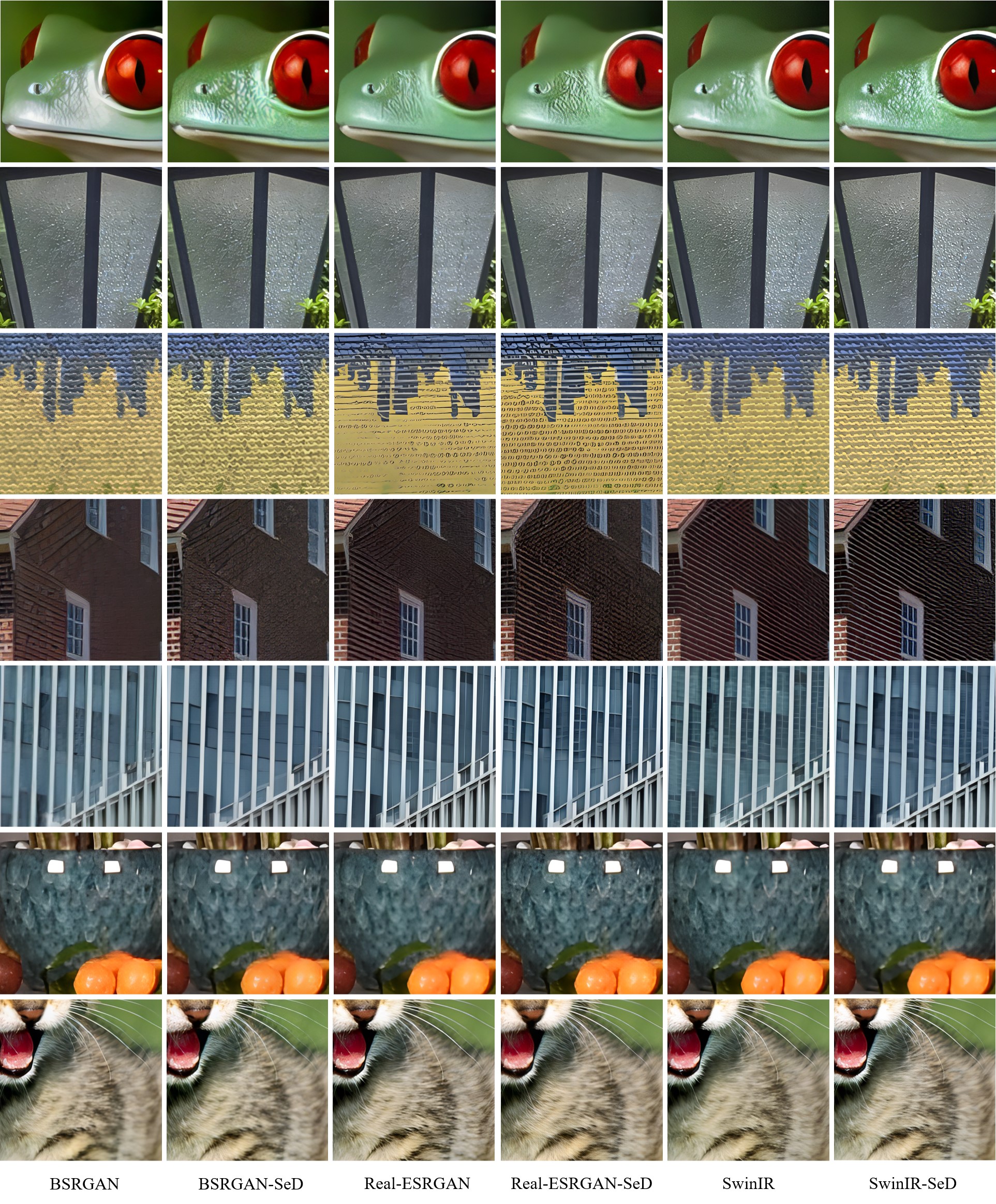}
    \caption{More visual comparisons between SeD and prominent GAN-based methods on real-world image SR of natural sceneries. Zoom in for better view.}
    \label{fig:real1}
\end{figure*}

\begin{figure*}
    \centering
    \begin{subfigure}{.33\linewidth}
        \centering
        \includegraphics[width=0.95\linewidth]{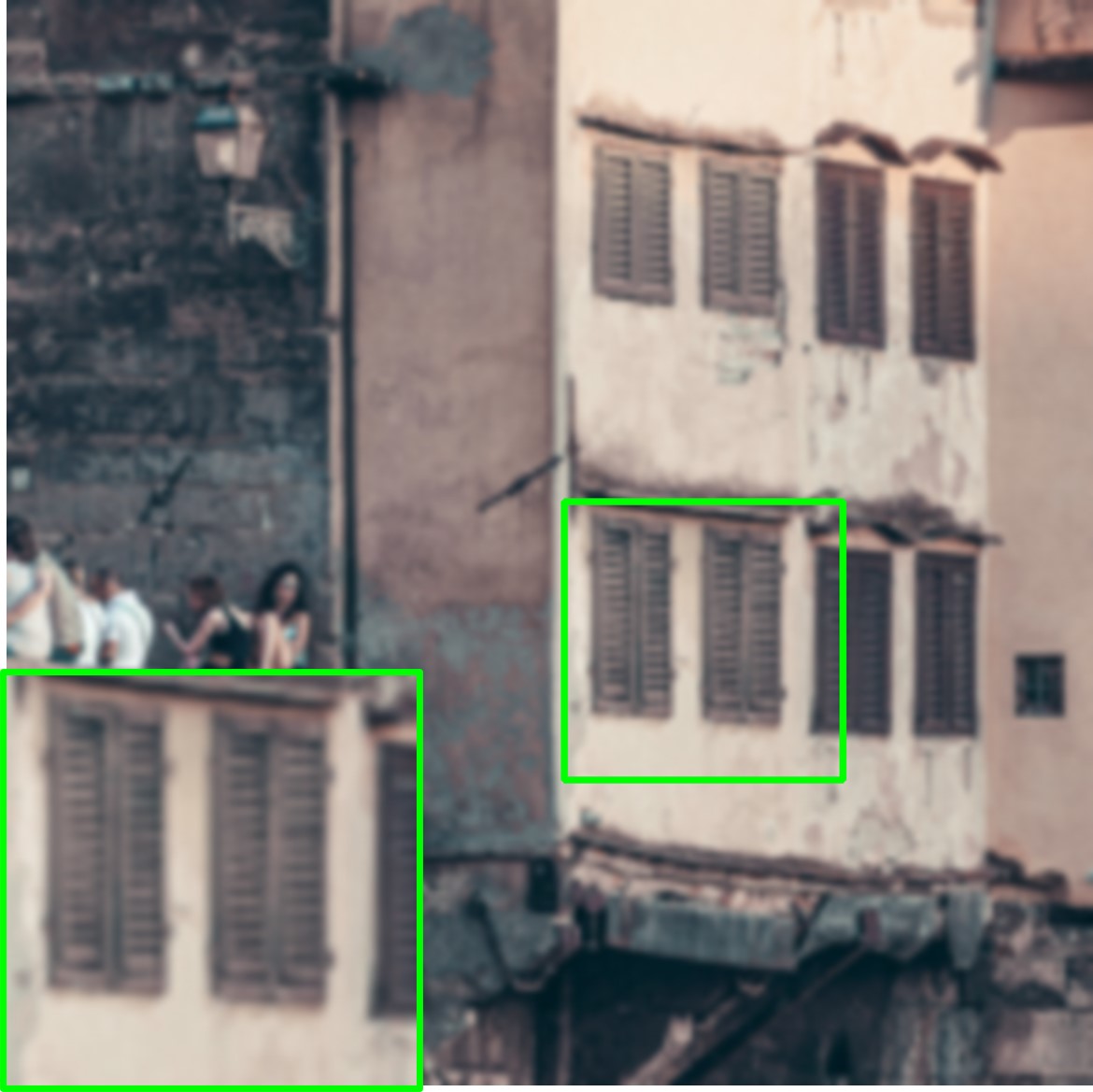}
        \caption*{Bicubic}
    \end{subfigure}%
    \begin{subfigure}{.33\linewidth}
        \centering
        \includegraphics[width=0.95\linewidth]{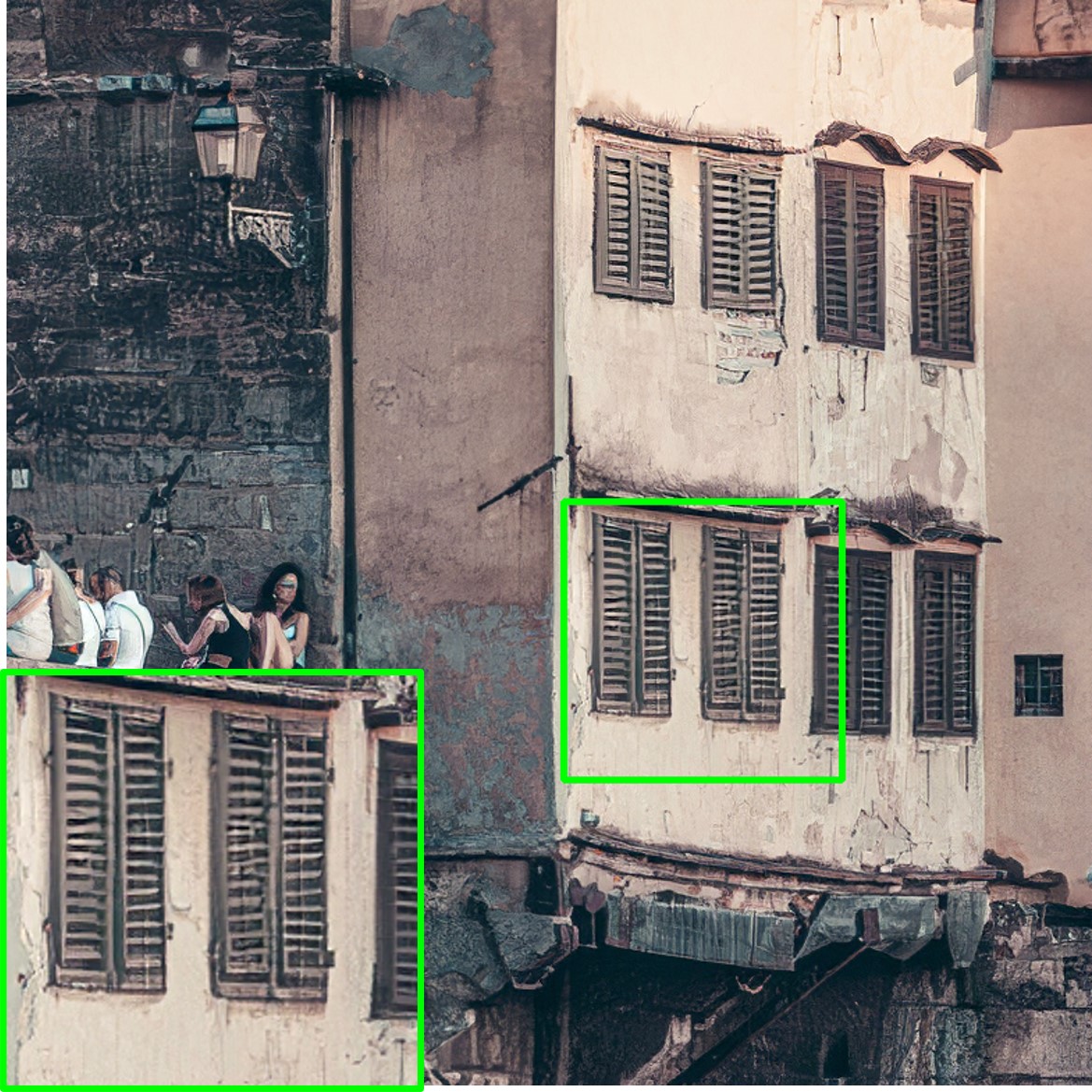}
        \caption*{ESRGAN~\cite{ESRGAN}}
    \end{subfigure}%
    \begin{subfigure}{.33\linewidth}
        \centering
        \includegraphics[width=0.95\linewidth]{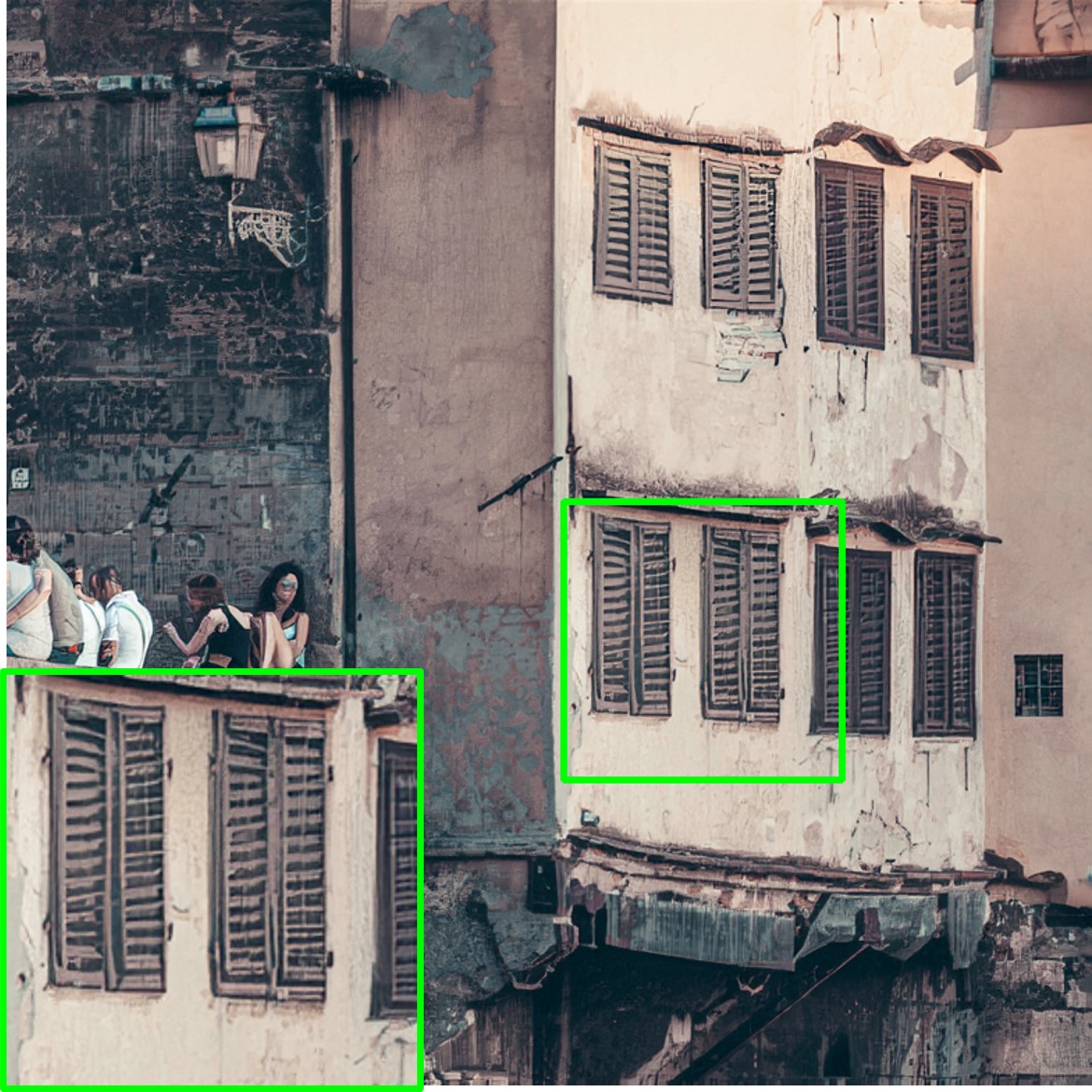}
        \caption*{USRGAN~\cite{USRNet}}
    \end{subfigure}
    \begin{subfigure}{.33\linewidth}
        \centering
        \includegraphics[width=0.95\linewidth]{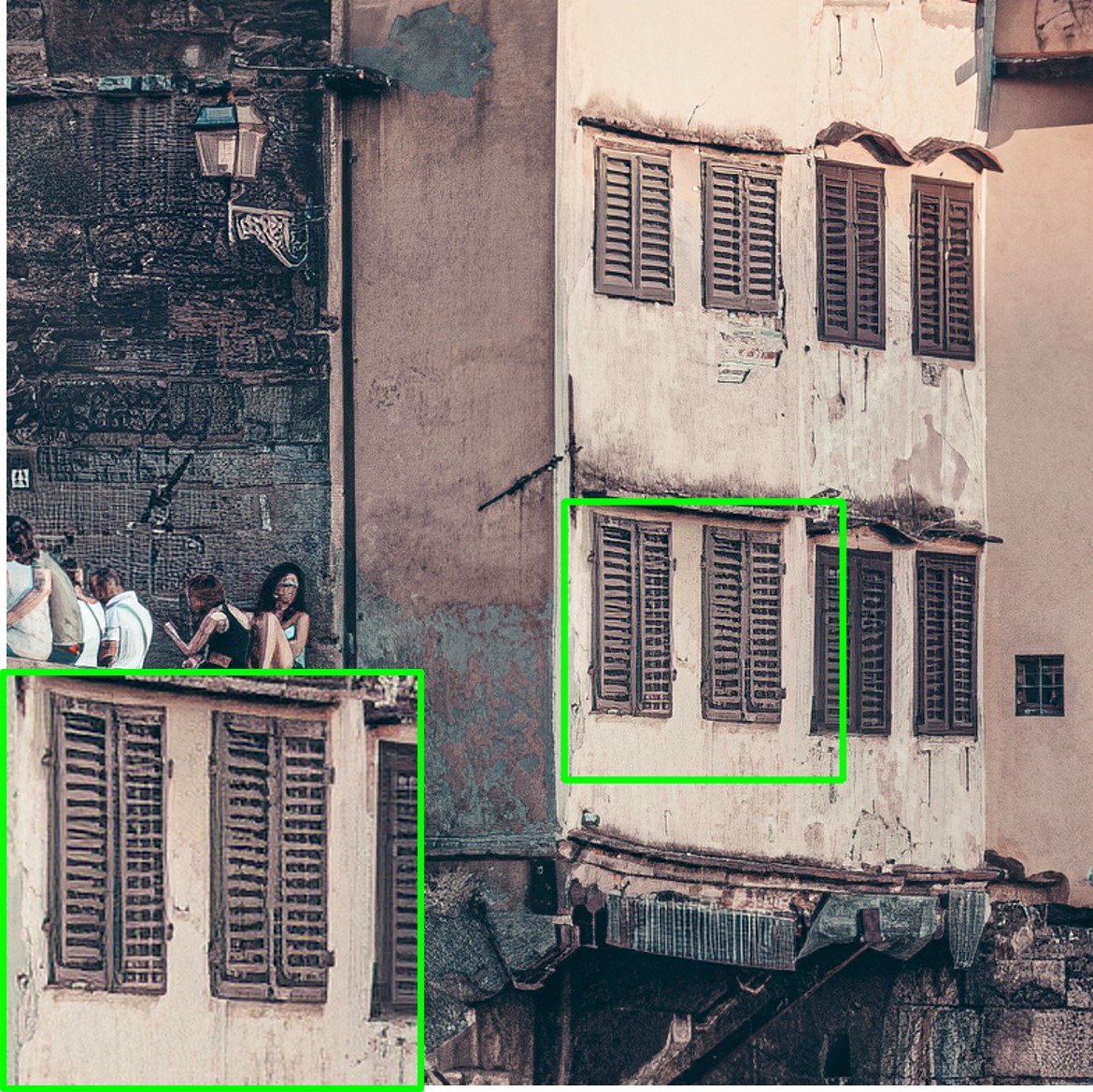}
        \caption*{RRDB+LDL~\cite{LDL}}
    \end{subfigure}%
    \begin{subfigure}{.33\linewidth}
        \centering
        \includegraphics[width=0.95\linewidth]{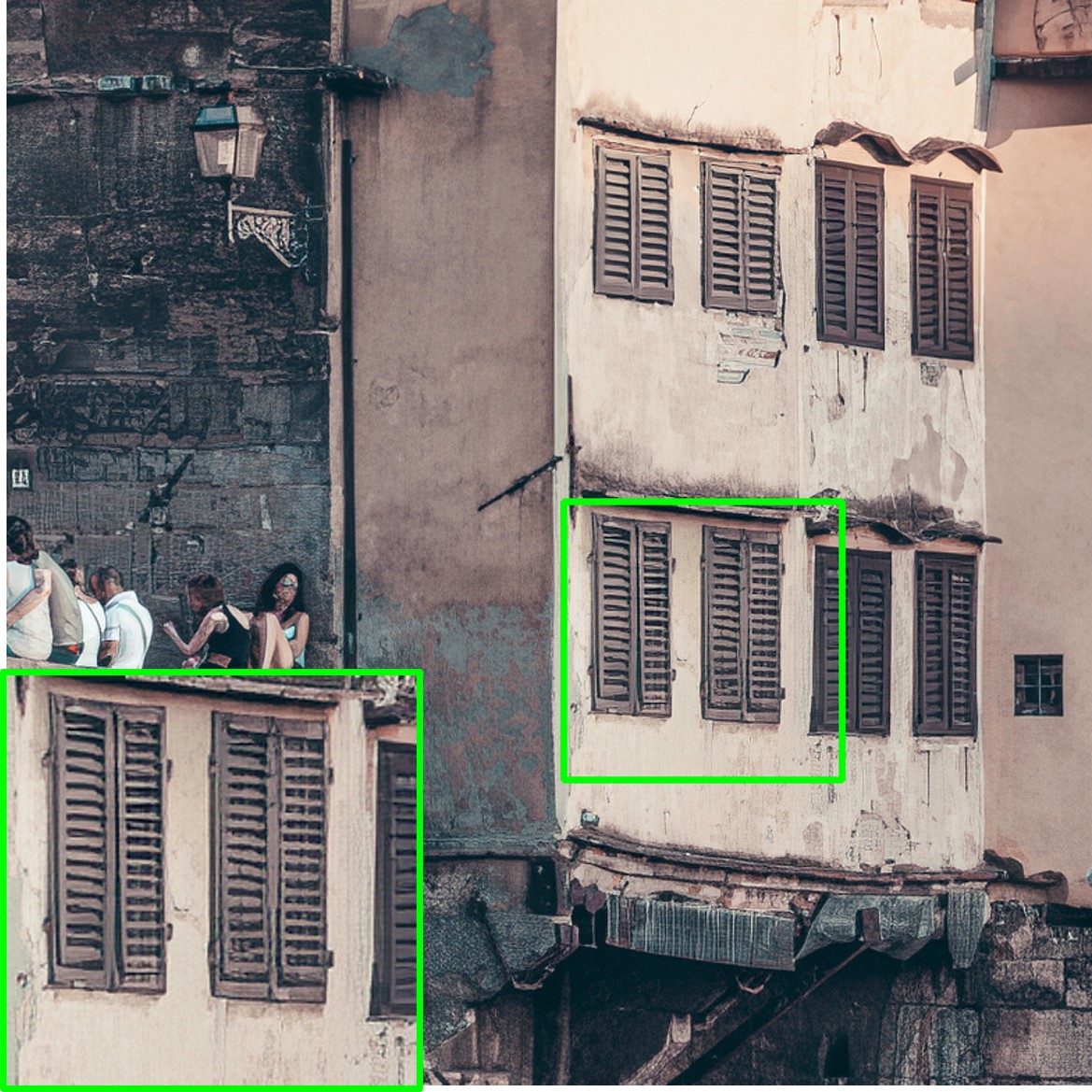}
        \caption*{RRDB+P+SeD}
    \end{subfigure}%
    \begin{subfigure}{.33\linewidth}
        \centering
        \includegraphics[width=0.95\linewidth]{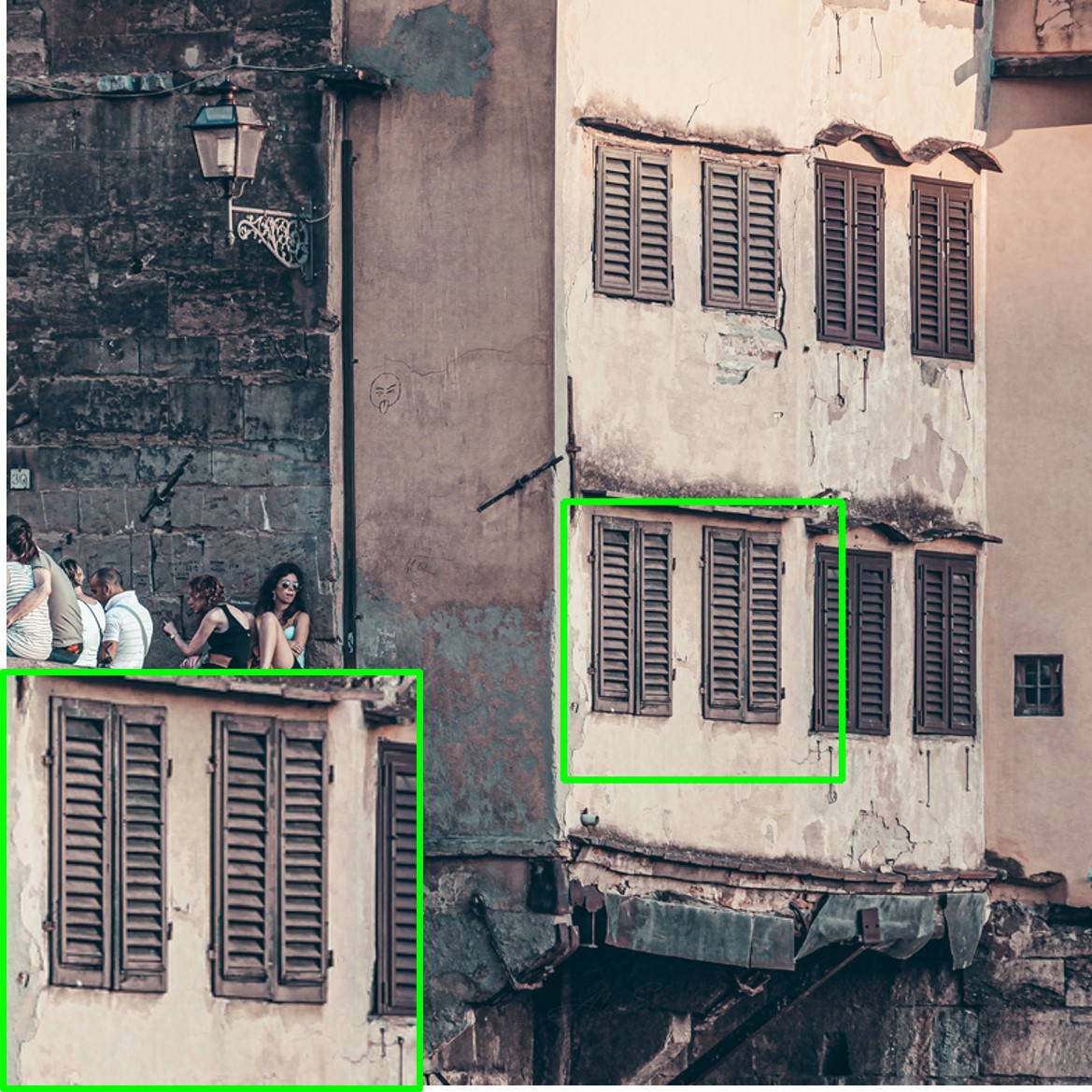}
        \caption*{HR}
    \end{subfigure}%
    
    \caption{Visual Comparisons between SeD and other methods on LSDIR~\cite{li2023lsdir} (img\_0000127 with resolution 780 $\times$ 780).}
    \label{lsdir1}
\end{figure*}

\begin{figure*}
    \flushleft
    \begin{subfigure}{.33\linewidth}
        \centering
        \includegraphics[width=0.95\linewidth]{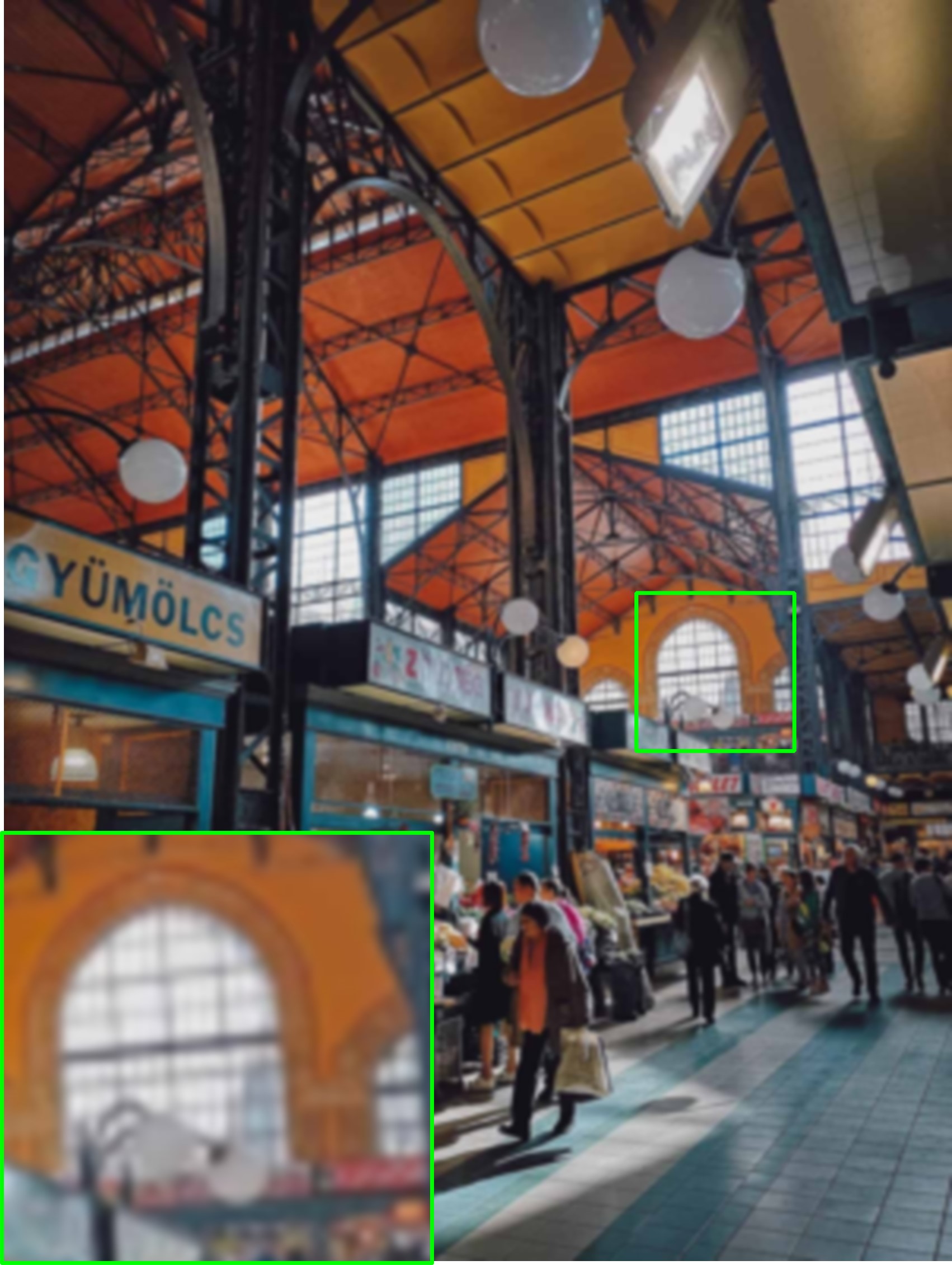}
        \caption*{Bicubic}
    \end{subfigure}%
    \begin{subfigure}{.33\linewidth}
        \centering
        \includegraphics[width=0.95\linewidth]{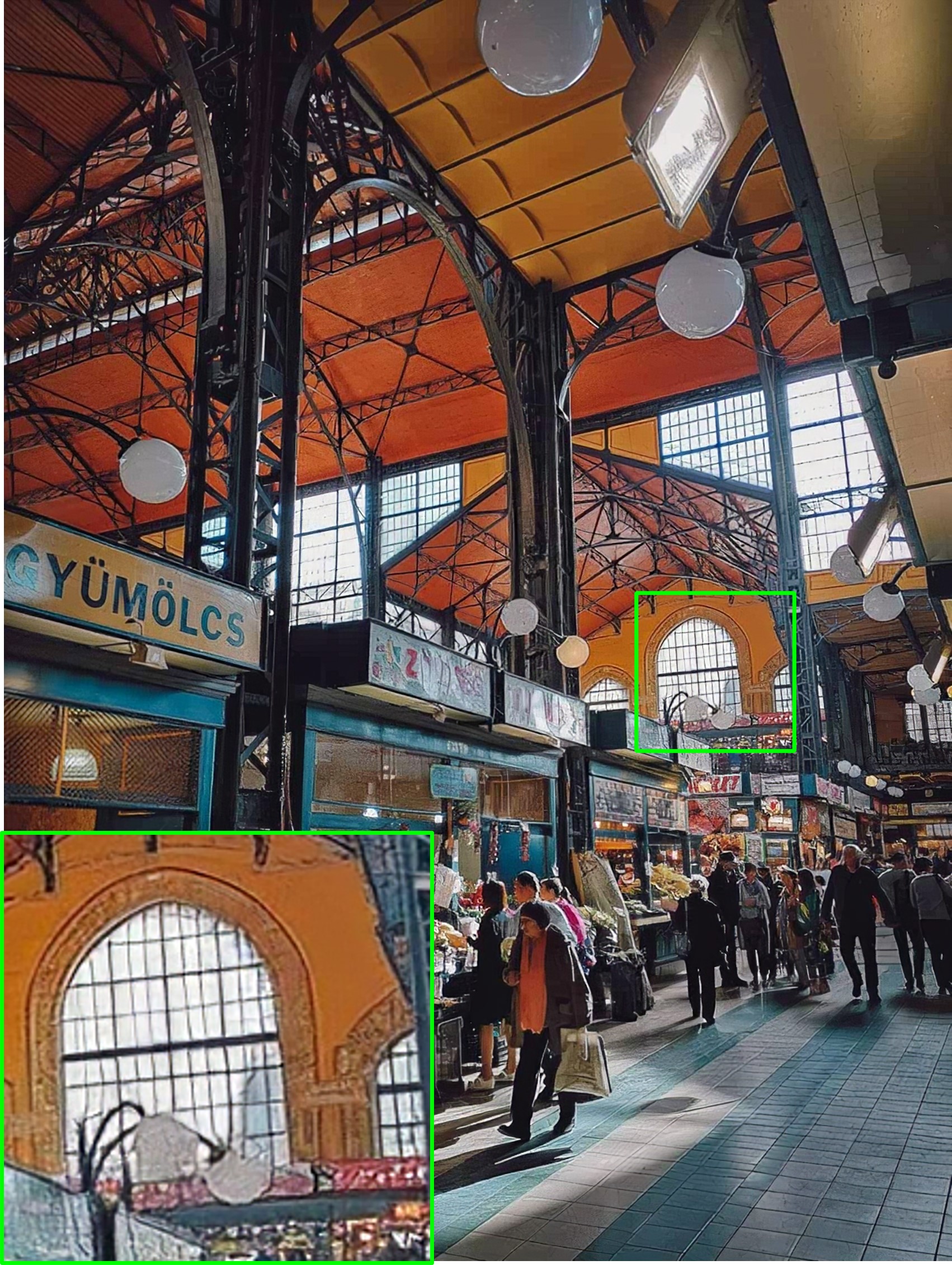}
        \caption*{ESRGAN~\cite{ESRGAN}}
    \end{subfigure}%
    \begin{subfigure}{.33\linewidth}
        \centering
        \includegraphics[width=0.95\linewidth]{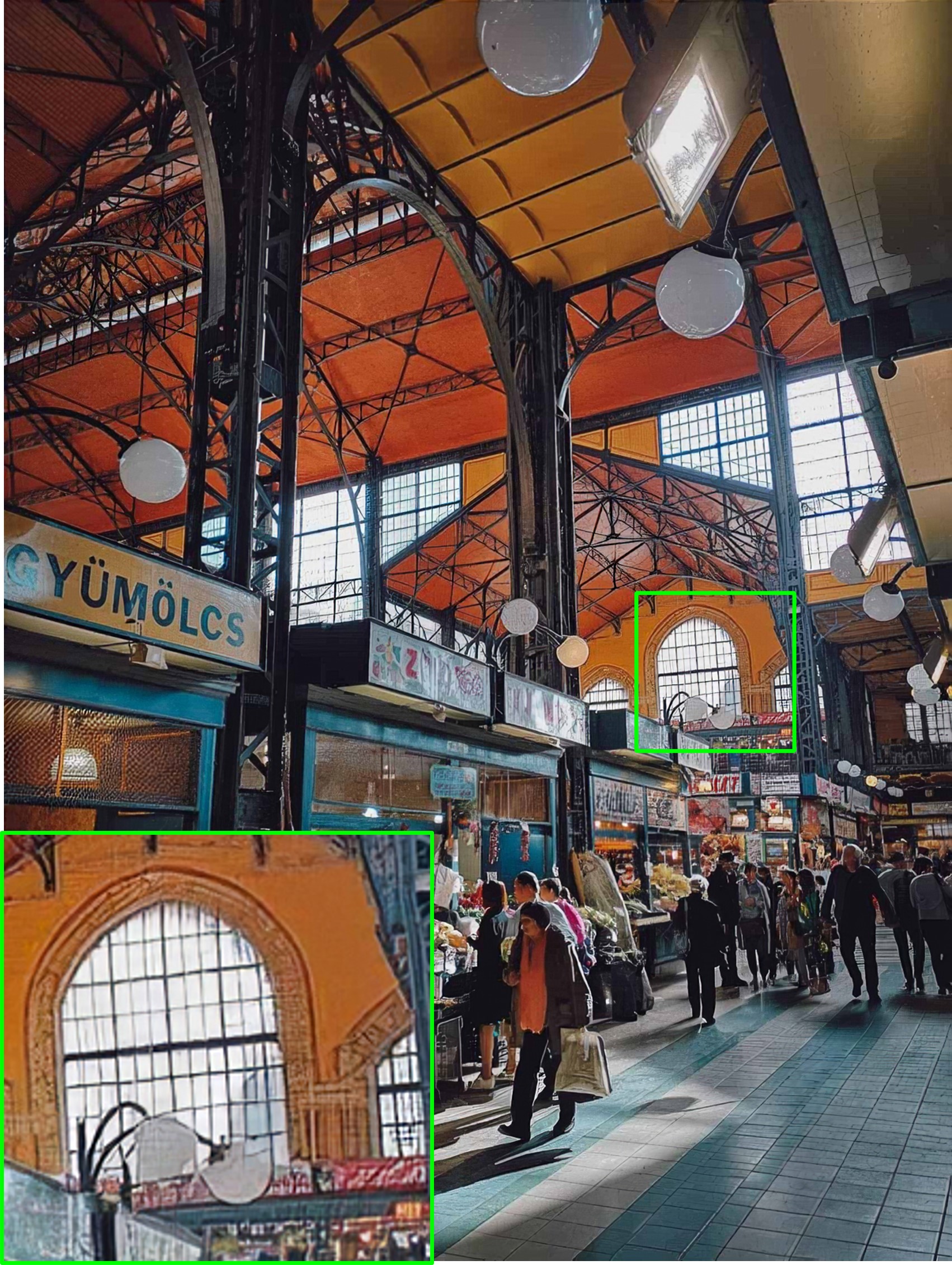}
        \caption*{USRGAN~\cite{USRNet}}
    \end{subfigure}
    \begin{subfigure}{.33\linewidth}
        \centering
        \includegraphics[width=0.95\linewidth]{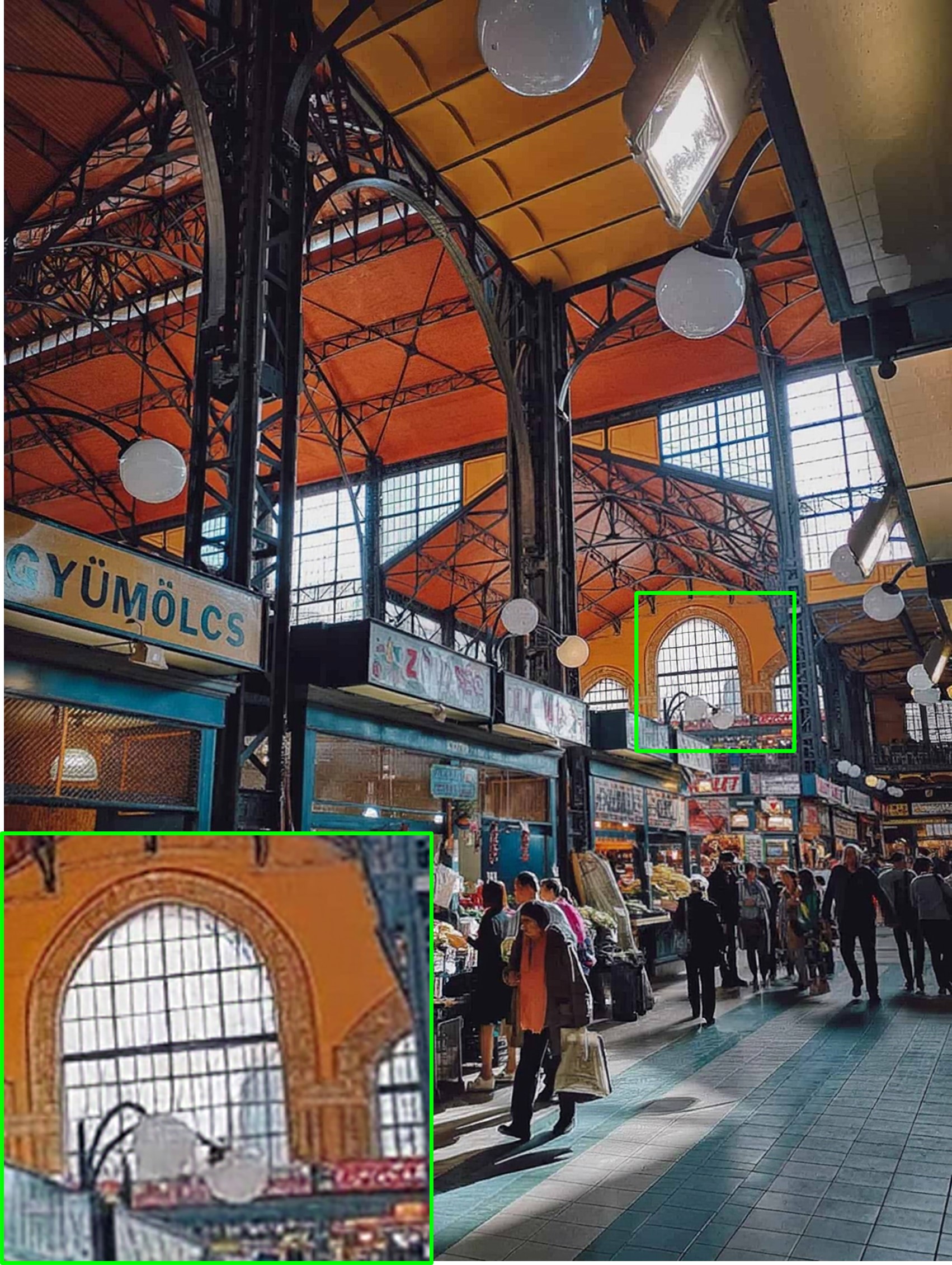}
        \caption*{RRDB+LDL~\cite{LDL}}
    \end{subfigure}%
    \begin{subfigure}{.33\linewidth}
        \centering
        \includegraphics[width=0.95\linewidth]{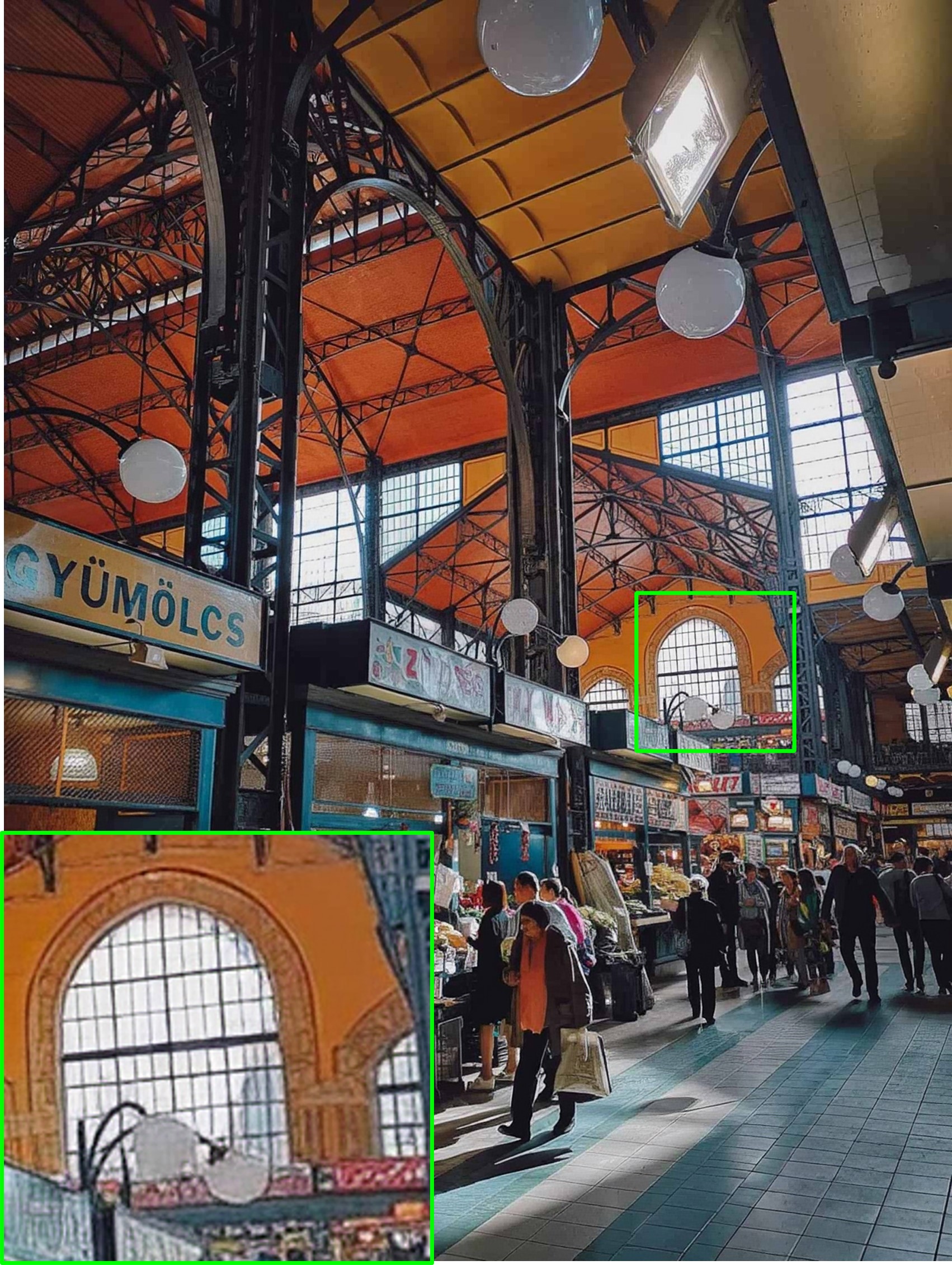}
        \caption*{RRDB+P+SeD}
    \end{subfigure}%
    \begin{subfigure}{.33\linewidth}
        \centering
        \includegraphics[width=0.95\linewidth]{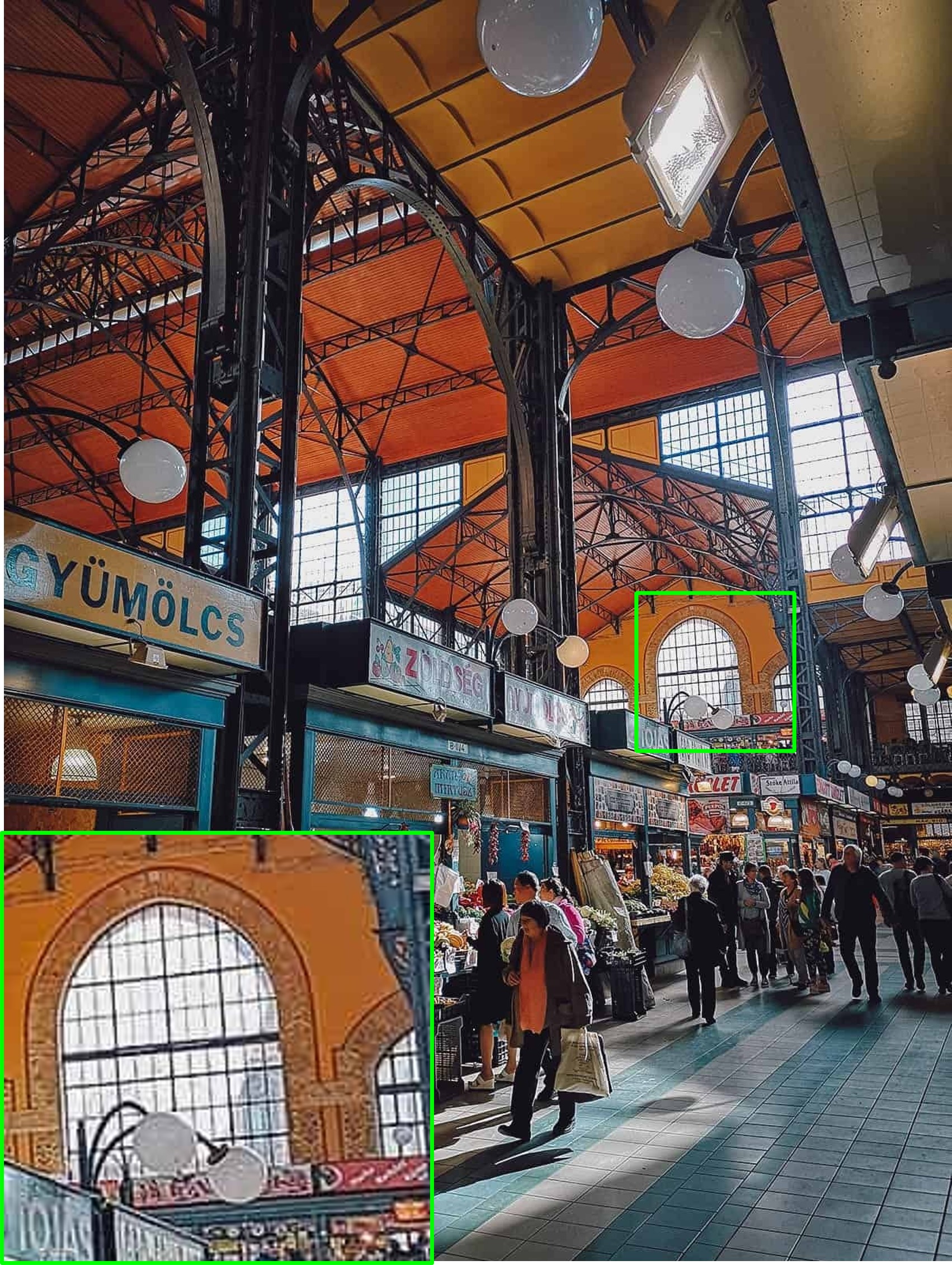}
        \caption*{HR}
    \end{subfigure}%
    
    \caption{Visual Comparisons between SeD and other methods on HQ-50K~\cite{yang2023hq} (complex/00020 with resolution 1200 $\times$ 1596).}
    \label{hq1}
\end{figure*}

\begin{figure*}
    \flushleft
    \begin{subfigure}{.50\linewidth}
        \centering
        \includegraphics[width=0.95\linewidth]{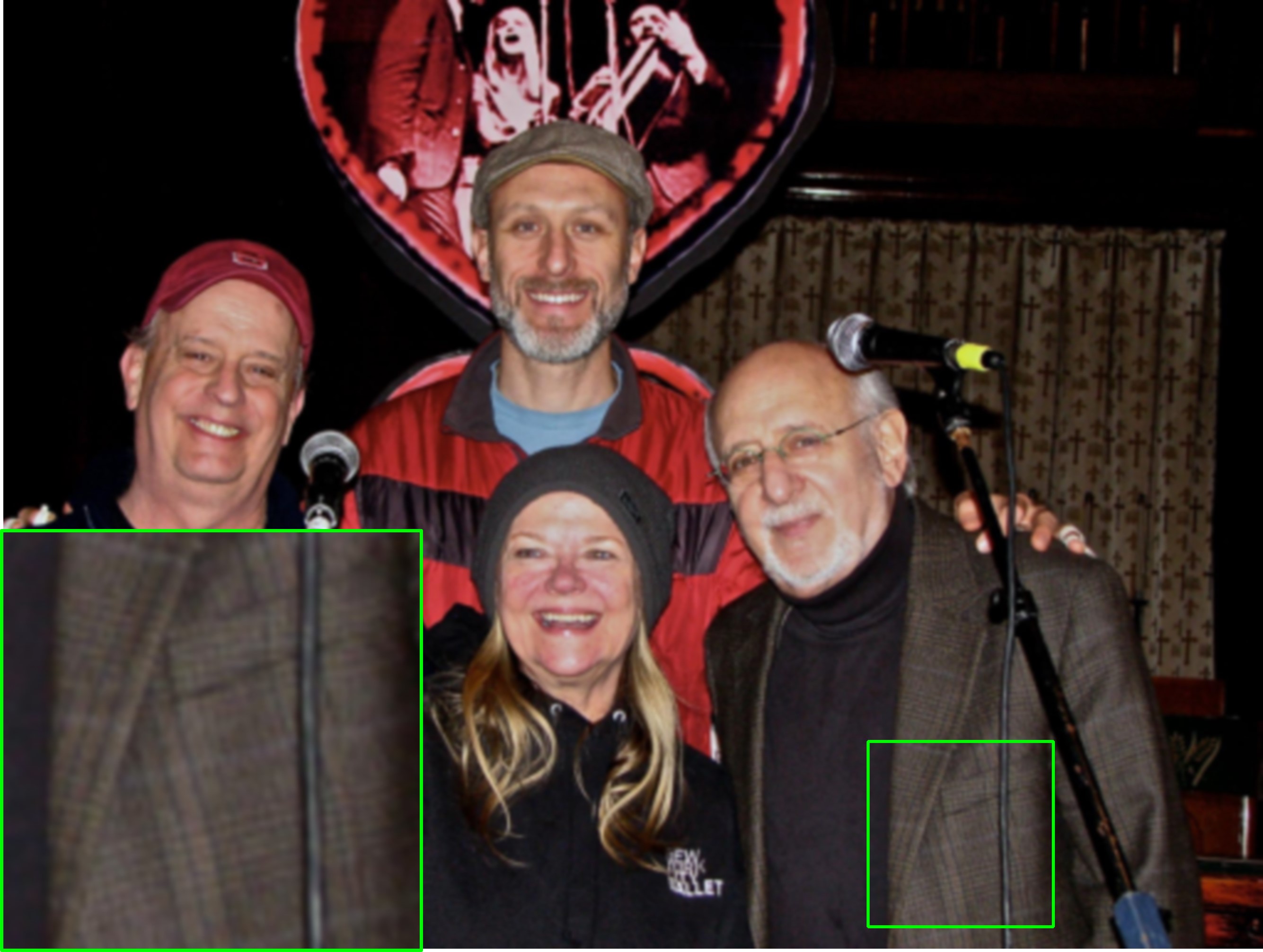}
        \caption*{Bicubic}
    \end{subfigure}%
    \begin{subfigure}{.50\linewidth}
        \centering
        \includegraphics[width=0.95\linewidth]{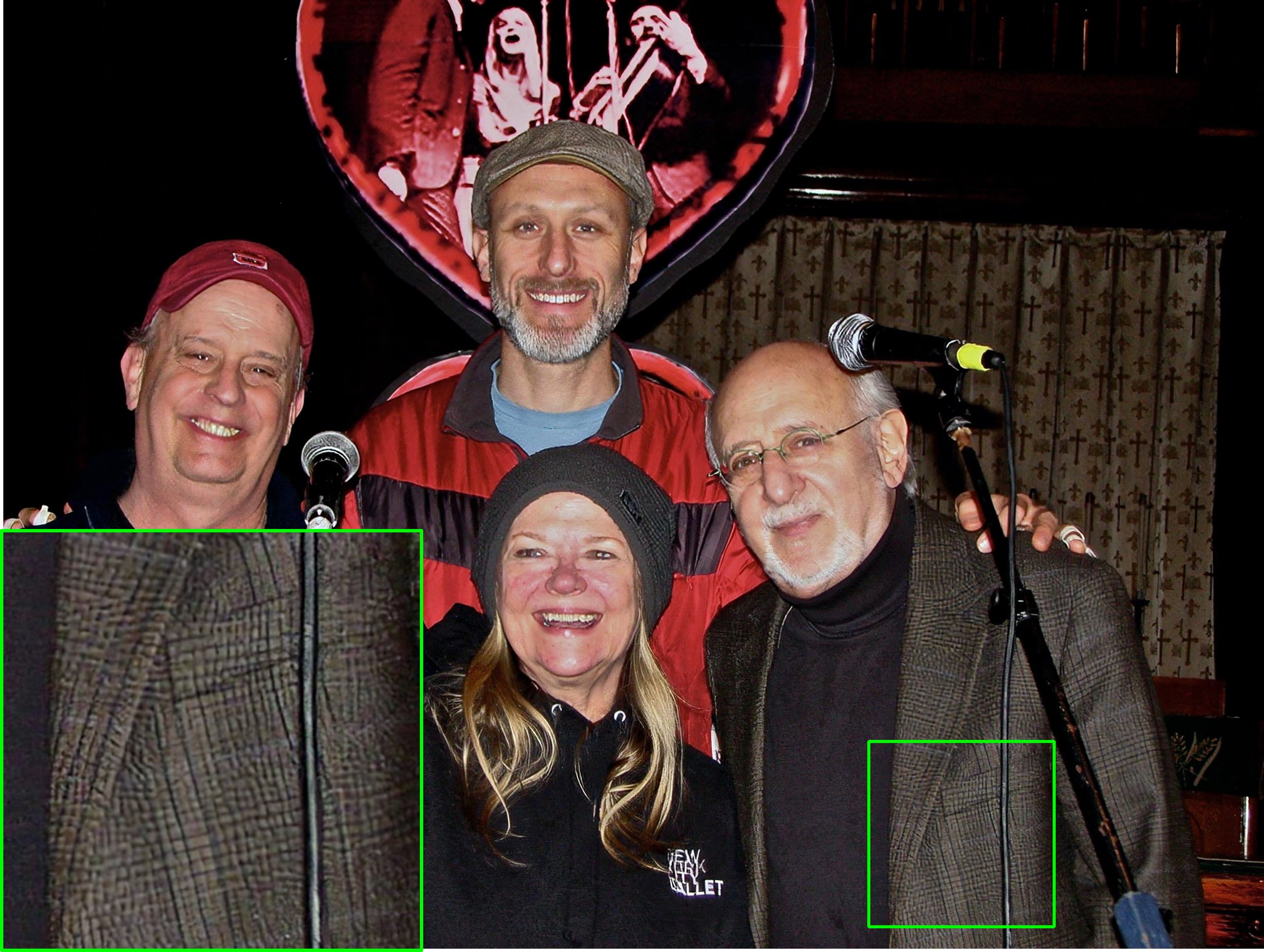}
        \caption*{ESRGAN~\cite{ESRGAN}}
    \end{subfigure}
    
    \begin{subfigure}{.50\linewidth}
        \centering
        \includegraphics[width=0.95\linewidth]{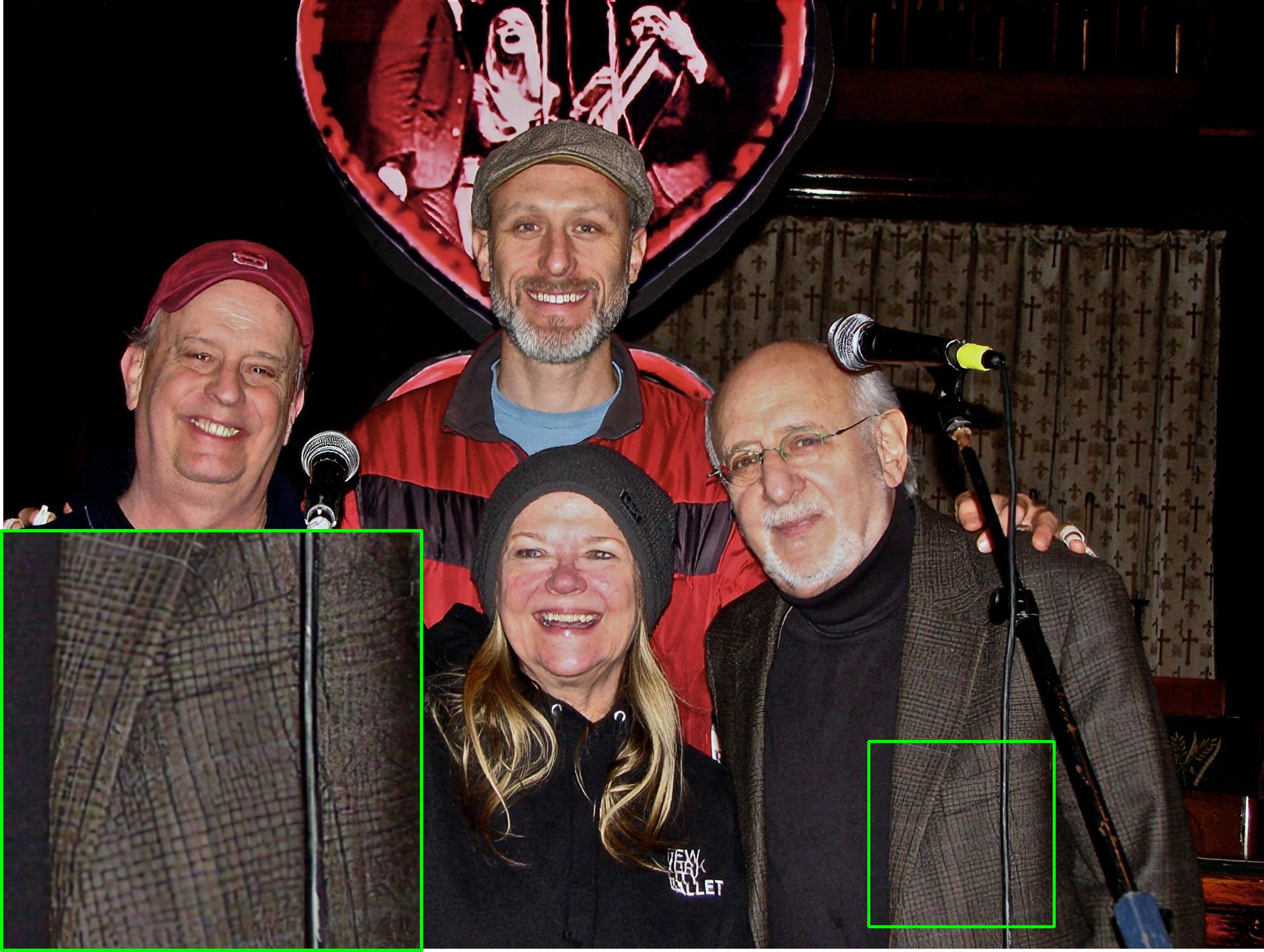}
        \caption*{USRGAN~\cite{USRNet}}
    \end{subfigure}%
    \begin{subfigure}{.50\linewidth}
        \centering
        \includegraphics[width=0.95\linewidth]{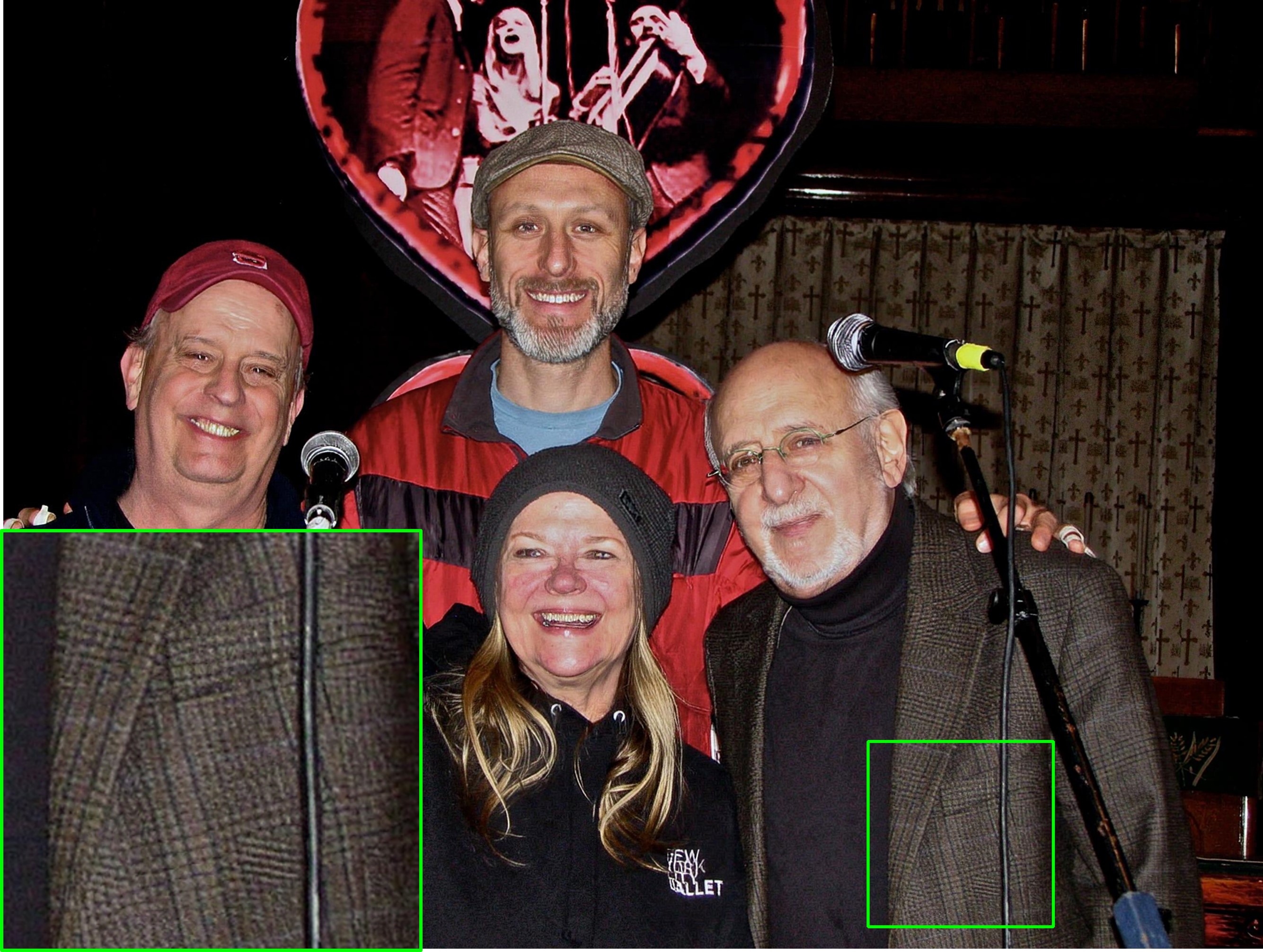}
        \caption*{RRDB+LDL~\cite{LDL}}
    \end{subfigure}
    \begin{subfigure}{.50\linewidth}
        \centering
        \includegraphics[width=0.95\linewidth]{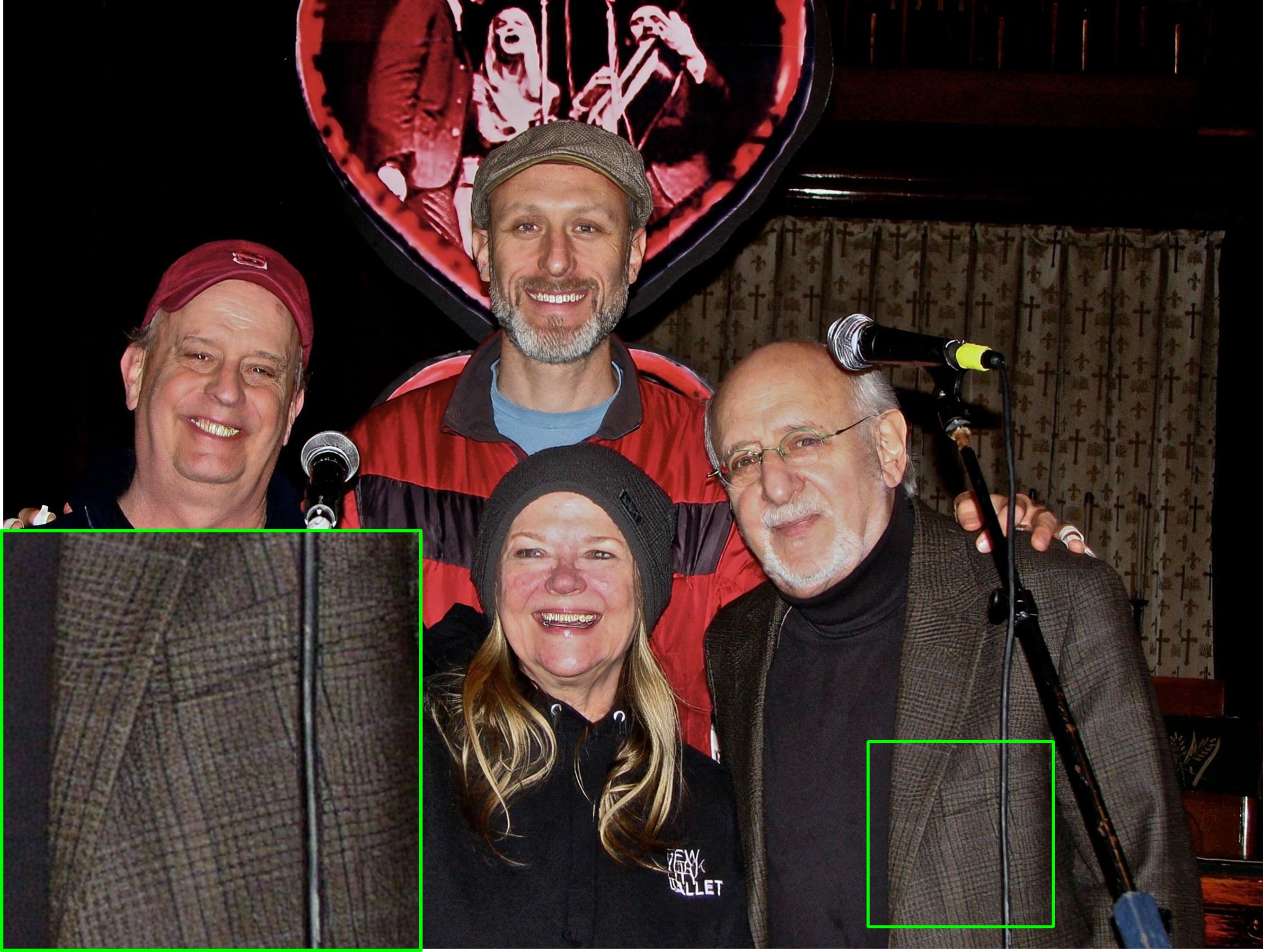}
        \caption*{RRDB+P+SeD}
    \end{subfigure}%
    \begin{subfigure}{.50\linewidth}
        \centering
        \includegraphics[width=0.95\linewidth]{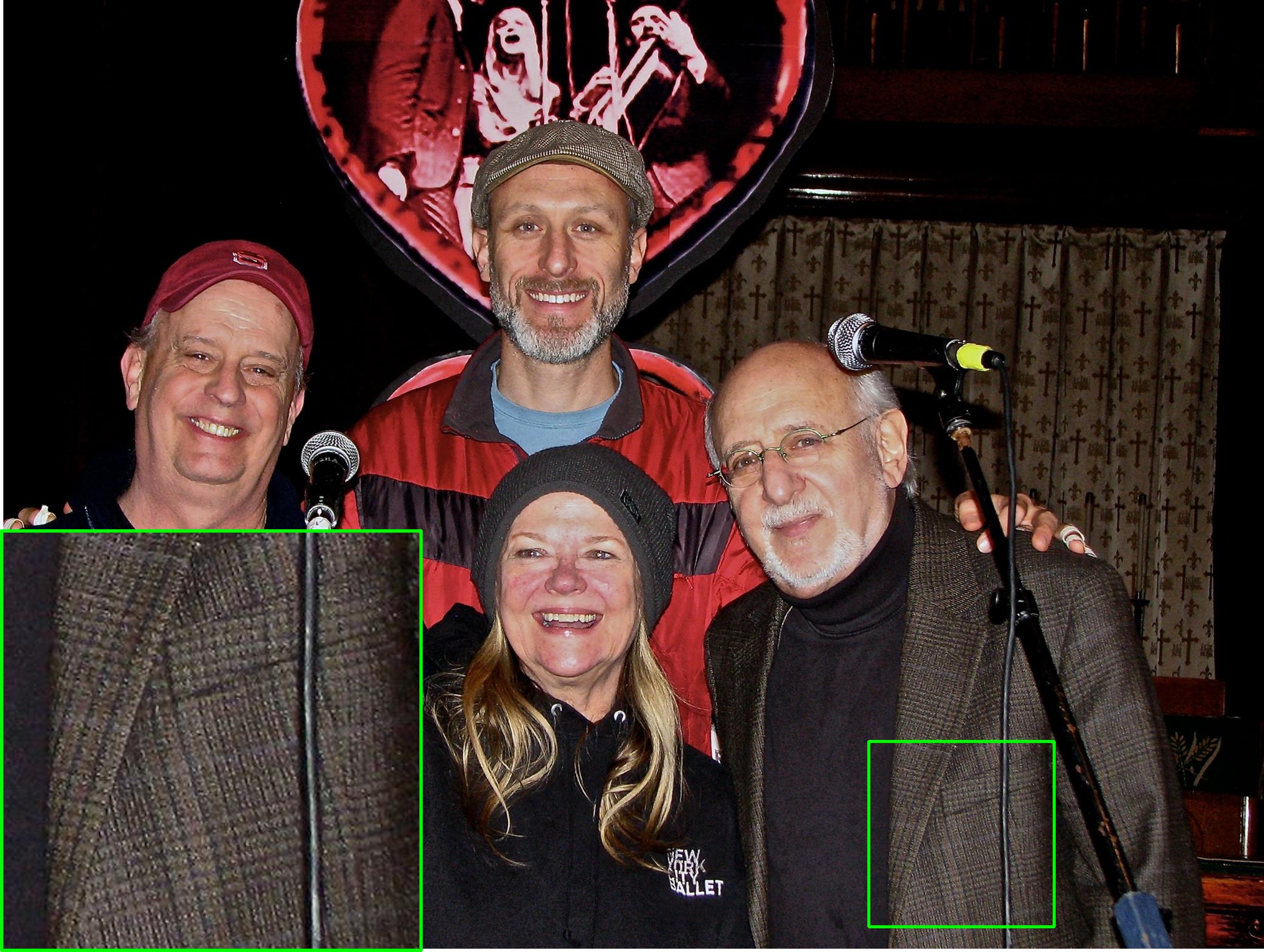}
        \caption*{HR}
    \end{subfigure}%
    
    \caption{Visual Comparisons between SeD and other methods on HQ-50K~\cite{yang2023hq} (people/00010 with resolution 2040 $\times$ 1536).}
    \label{hq2}
\end{figure*}

Then, we show more qualitative comparisons with previous GAN-based SR methods in classical and real-world image SR. As shown in Fig.~\ref{fig:classical2} and Fig.~\ref{fig:real1}, our SeD enables SR networks to correctly restore repeating textures, where previous works typically fail to deal with. Moreover, with semantic guidance, our SeD is capable of recovering more fine-grained textures (\eg, twigs, furs and windows) in natural sceneries, in the meanwhile reducing the artifacts of super-resolved images. These conclusions remain valid for large-scale benchmarks, as illustrated in Fig.~\ref{lsdir1}, Fig.~\ref{hq1} and Fig.~\ref{hq2}, which further demonstrates the effectiveness of our SeD.

\end{document}